\documentclass[english,twocolumn,showpacs,preprintnumbers,amsmath,amssymb]{revtex4}
\usepackage[T1]{fontenc}
\usepackage[latin1]{inputenc}
\usepackage{graphicx}
\usepackage{amssymb}

\makeatletter

\makeatletter

\makeatletter

\usepackage{colordvi}

\usepackage{color}

\makeatletter

\usepackage{dcolumn}

\usepackage{bm}

\makeatother

\makeatother

\makeatother

\usepackage{babel}
\makeatother
\begin{document}

\title{Development of Fractal Geometry in a 1+1 Dimensional Universe}

\author{Bruce N. Miller}

\affiliation{Department of Physics and Astronomy, Texas Christian University,
\\
 Fort Worth, Texas 76129}

\email{b.miller@tcu.edu}

\homepage{http://personal.tcu.edu/ bmiller}

\author{Jean-Louis Rouet}

\affiliation{Institut des Sciences de la Terre d'Orl\'{e}ans (ISTO) - UMR 6113
CNRS/Universit\'{e} d'Orl\'{e}ans,\\
 1A, rue de la F\'{e}rollerie and Laboratoire de Math\'{e}matique,
Applications et Physique Math\'{e}matique - UMR 6628 CNRS/Universit\'{e}
d'Orl\'{e}ans,\\
 UFR des Sciences, F-45067 Orl\'{e}ans Cedex 2, France}

\author{Emmanuel Le Guirrec}

\affiliation{Laboratoire de Math\'{e}matique, Applications et Physique Math\'{e}matique
- UMR 6628 CNRS/Universit\'{e} d'Orl\'{e}ans,\\
 UFR des Sciences, F-45067 Orl\'{e}ans Cedex 2, France\\
 }

\date{\today}

\begin{abstract}
Observations of galaxies over large distances reveal the possibility
of a fractal distribution of their positions. The source of fractal
behavior is the lack of a length scale in the two body gravitational
interaction. However, even with new, larger, sample sizes from recent
surveys, it is difficult to extract information concerning fractal
properties with confidence. Similarly, simulations with a billion
particles only provide a thousand particles per dimension, far too
small for accurate conclusions. With one dimensional models these
limitations can be overcome by carrying out simulations with on the
order of a quarter of a million particles without compromising the
computation of the gravitational force. Here the multifractal properties
of a group of these models that incorporate different features of
the dynamical equations governing the evolution of a matter dominated
universe are compared. The results share important similarities with
galaxy observations, such as hierarchical clustering and apparent
bifractal geometry. They also provide insights concerning possible
constraints on length and time scales for fractal structure. They
clearly demonstrate that fractal geometry evolves in the $\mu$ (position,
velocity) space. The observed properties are simply a shadow (projection)
of higher dimensional structure. 
\end{abstract}



\maketitle

\section{Introduction}

Starting with the work of Vaucoulours,\cite{Vau} questions have been
posed about the geometric properties of the distribution of matter
in the universe. Necessary for all modern cosmologies are the assumptions
of homogeneity and isotropy of the mass distribution on large length
scales.\cite{peebles2} However, observations have shown the existence
of very large structures such as super-clusters and voids.\cite{peck}
Moreover, as technology has advanced, so has the length scale of the
largest observed structures. (For a recent review see Jones et. al.\cite{cosmo-rev}.)
It was proposed by Mandelbrot,\cite{man} based on work of Peebles,\cite{peebles2}
that the matter distribution in the universe is fractal. Support for
this conjecture came primarily from the computation of the pair correlation
function for the positions of galaxies, as well as from direct observation.
The fact that the correlation function was well represented by a power
law in the intergalactic distance seemed to support the fractal conjecture.
In fact, in the past it has been argued by Pietronero and colleagues
that the universe may even be fractal on all scales.\cite{cosmo-rev}\cite{Pietro_rev}
Of course, if this were true it would wreak havoc with the conclusions
of cosmological theory.

McCauley has looked at this issue from a few different perspectives.
He points out that since the power law behavior of the correlation
function is only quantitatively correct over a finite length scale,
the universe could not be a simple (mono) fractal.\cite{Mac} Then
the logical next question is whether or not the matter distribution
is multifractal. By considering counts in cells, Bailin and Schaeffer
hypothesized some time ago that the distribution of matter is approximately
bifractal, i.e. a superposition of fractals with two distinct scaling
laws.\cite{Bal} McCauley has also addressed the possibility of inhomogeneity.
Working with Martin Kerscher, he investigated whether the universe
has multifractal geometry. His approach was to examine the point-wise
dimension \cite{Ott} by looking for local scaling of the density
around individual galaxies in two catalogues. Their conclusion was
that, even if local scaling were present, there are large fluctuations
in the scaling law. Moreover, the sample sizes were not sufficient
to be able to extract good local scaling exponents. Although larger
galaxy catalogues have become available since their analysis, they
do not meet the restrictive criteria of sufficient size, as well as
uniform extension in all directions, necessary to measure local dimensions.
\cite{cosmo-rev} Their final conclusion was that, while the geometry
of the observed universe is certainly not monofractal, it was not
possible to irrefutably conclude whether it is multifractal, or whether
the assumptions of homogeneity and isotropy prevail at large length
scales.\cite{Mac}

If current observations aren't able to let us determine the geometry
of the universe, then we need to turn to theory and simulation. For
the most part, we are concerned with the evolution of matter after
the period of recombination. Consequently the Hubble expansion has
slowed down sufficiently that Newtonian dynamics is applicable, at
least within a finite region of space.\cite{Newtap} A number of theoretical
approaches to the computation of fractal dimensions have been investigated.
Each of them is predicated on some external assumption which is not
yet verified. For example, de Vega et. al. assume that the universe
is close to a thermodynamic critical point \cite{vega1} , while Grujic
explores a field theory where vacuum energy predominates over inert
matter, and the latter is assumed to have a fractal structuring \cite{Grujic1}.
An examination of the recent literature reveals that theory has not
converged on a compelling and uniformly accepted theory of fractal
structure in the universe.

In the last few decades dynamical N-body simulation of cold dark matter
CDM has experienced rapid advances due to improvements in both algorithms
and technology.\cite{Bag,Ber} It is now possible to carry out gravitational
N-body simulations with upwards of $10^{9}$ point mass particles.
However, in order to employ simulation methods for systems evolving
over cosmological time, it is necessary to compromise the representation
of the gravitational interaction over both long and short length scales.
For example, tree methods are frequently employed for large separations,
typically the gravitational field is computed from a grid, and a short
range cut-off is employed to control the singularity in the Newtonian
pair potential.\cite{Bag,Ber} Unfortunately, even if the simulations
were perfect, a system of even $10^{9}$ particles provides only $10^{3}$
particles per dimension and would thus be insufficient to investigate
the fractal geometry with confidence.

As a logical consequence of these difficulties it was natural that
physicists would look to lower dimensional models for insight. Although
this sacrifices the correct dynamics, it provides an arena where accurate
computations with large numbers of particles can be carried out for
significant cosmological time. It is hoped that insights gained from
making this trade-off are beneficial. In one dimension, Newtonian
gravity corresponds to a system of infinitesimally thin, parallel,
mass sheets of infinite spatial extent. Since there is no curvature
in a 1+1 dimensional gravitational system, we cannot expect to obtain
equations of motion directly from general relativity. \cite{Mann1}
Then a question arises concerning the inclusion of the Hubble flow
into the dynamical formulation. This has been addressed in two ways:
In the earliest, carried out by Rouet and Feix, the scale function
was directly inserted into the one dimensional dynamics.\cite{Rouet1,Rouet2}
Alternatively, starting with the usual three dimensional equations
of motion and embedding a stratified mass distribution, Fanelli and
Aurell obtained a similar set of coupled differential equations for
the evolution of the system in phase space.\cite{quintic} While the
approaches are different, from the standpoint of mathematics the two
models are very similar and differ only in the values of a single
fixed parameter, the effective friction constant. Following Fanelli,\cite{quintic}
we refer to the former as the RF model and the latter as the Quintic,
or Q, model.

By introducing scaling in both position and time, in each model autonomous
equations of motion are obtained in the comoving frame. In addition
to the contribution for the gravitational field, there is now a background
term, corresponding to a constant negative mass distribution, and
a linear friction term. By eliminating the friction term a Hamiltonian
version can also be constructed. At least three other one dimensional
models have also been investigated, one consisting of Newtonian mass
sheets which stick together whenever they cross,\cite{sticky} one
evolved by directly integrating the Zeldovich equations,\cite{Zel,Tat}
and the continuous system satisfying Burger's equation \cite{Bur}.
In addition, fractal behavior has been studied in the autonomous one
dimensional system where there is no background Hubble flow.\cite{KK1,KK2}

In dynamical simulations, both the RF and Quintic model clearly manifest
the development of hierarchical clustering in both configuration and
$\mu$ space (the projection of the phase space on the position-velocity
plane). In common with the observation of galaxy positions, as time
evolves both dense clusters and relatively empty regions (voids) develop.
In their seminal work, by computing the box counting dimension for
the RF model, Rouet and Feix were able to directly demonstrate the
formation of fractal structure.\cite{Rouet1,Rouet2} They found a
value of about 0.6 for the box counting dimension of the well evolved
mass points in the configuration space, indicating the formation of
a robust fractal geometry. In a later work, Miller and Rouet investigated
the generalized dimension of the RF model.\cite{MRex} In common with
the analysis of galaxy observations by Bailin and Schaeffer \cite{Bal}
they found evidence for bi-fractal geometry. Although they did not
compute actual dimensions, later Fanelli et. al. also found a suggestion
of bi-fractal behavior in the model without friction \cite{Aurell}.
It is then not surprising that the autonomous, isolated, gravitational
system which does not incorporate the Hubble flow also manifests fractal
behavior for short times as long as the virial ratio realized by the
initial conditions is very small.\cite{KK1,KK2} In addition, in the
one dimensional model of turbulence governed by Burger's equation,
the formation of shocks has the appearance of density singularities
that are similar to the clusters found in the RF and Quintic model.
A type of bi-fractal geometry has also been demonstrated for this
system.\cite{Bur}

Below we present the results of our recent investigation of multifractal
properties of the Quintic model and the model without friction. In
section II we will first describe the systems. Then we will give a
straightforward derivation of the equations of motion and explain
how they differ from the other models mentioned above. In section
III we will explain how the simulations were carried out and describe
their qualitative features. In section IV we define the generalized
dimension and other fractal measures and present our approach for
computing them. In section V we will present the results of the multifractal
analysis. Finally conclusions will be presented in section VI.

\section{Description of the System}

We consider a one dimensional gravitational system (OGS) of $N$ parallel,
infinitesimally thin, mass sheets each with mass per unit area $m$
. The sheets are constrained to move perpendicular to their surface.
Therefore we can construct a Cartesian axis $x$ perpendicular to
the $N$ sheets. We imagine that the sheets are labeled by $j=1,...,N$.
From its intersection with the $x$-axis, each sheet can be assigned
the coordinate $x_{j}$ and the corresponding velocity $v_{j}$. From
Gauss' law, the force on each sheet is a constant proportional to
the net difference between the total \char`\"{}mass\char`\"{} (really
surface density) to its right and left. The equations of motion are

\begin{equation}
\frac{dx_{i}}{dt}=v_{i},\quad\frac{dv_{i}}{dt}=E_{i}=2\pi mG[N_{R,i}-N_{L,i}]\label{1d}\end{equation}
 where $E_{i}$ is the gravitational field experienced by the i$^{th}$
particle and $N_{R,i}$ ($N_{L,i}$) is the number of particles (sheets)
to its right (left) . Because the acceleration of each particle is
constant between crossings, the equations of motion can be integrated
exactly. Therefore it is not necessary to use numerical integration
to follow the trajectory in phase space. As a consequence it was possible
to study the system dynamics on the earliest computers and it can
be considered the gravitational analogue of the Fermi-Pasta-Ulam system.\cite{FPU}
It was first employed by Lecar and Cohen to investigate the relaxation
of an $N$ body gravitational system.\cite{CL1} Although it was first
thought that the $N$ body system would reach equilibrium with a relaxation
time proportional to $N^{2}$ ,\cite{hohl} this was not born out
by simulations.\cite{WMS} Partially because of its reluctance to
reach equilibrium, both single and two component versions of the system
have been studied exhaustively in recent years.\cite{yawn2,yawn3}\cite{poshthirring}\cite{MPT1}\cite{MPT2}\cite{Rouet3}
Most recently it has been demonstrated that, for short times and special
initial conditions, the system evolution can be modeled by an exactly
integrable system.\cite{nearby}

To construct and explore a cosmological version of the OGS we introduce
the scale factor $A(t)$ for a matter dominated universe.\cite{peebles2}
Moreover, as discussed in the introduction, we are interested in the
development of density fluctuations following the time of recombination
so that electromagnetic forces can be ignored. For that, and later,
epochs, the Hubble expansion has slowed sufficiently that Newtonian
dynamics provides an adequate representation of the motion in a finite
region.\cite{Newtap} Then, in a $3+1$ dimensional universe, the
Newtonian equations governing a mass point are simply

\begin{equation}
\frac{d\mathbf{r}}{dt}=\mathbf{v,\qquad}\frac{d\mathbf{v}}{dt}=\mathbf{E}_{g}\mathbf{(r,}t\mathbf{)}\label{3d}\end{equation}
 where, here, $\mathbf{E}_{g}\mathbf{(r,}t\mathbf{)\ }$is the gravitational
field. We wish to follow the motion in a frame of reference where
the average density remains constant, i.e. the co-moving frame. Therefore
we transform to a new space coordinate which scales the distance according
to $A(t)$. Writing $\mathbf{r}=A(t)\mathbf{x}$ we obtain

\begin{equation}
\frac{d^{2}\mathbf{x}}{dt^{2}}+\frac{2}{A}\frac{dA}{dt}\frac{d\mathbf{x}}{dt}+\frac{1}{A}\frac{d^{2}A}{dt^{2}}\mathbf{x=}\frac{1}{A^{3}}\mathbf{E}_{g}(\mathbf{x},t)\label{scaldis}\end{equation}
 where, in the above we have taken advantage of the inverse square
dependence of the gravitational field to write $\mathbf{E}_{g}(\mathbf{x},t)=\frac{1}{A^{2}}\mathbf{E}_{g}\mathbf{(r},t)$
where the functional dependence is preserved. In a matter dominated
(Einstein deSitter) universe we find that

\begin{equation}
A(t)=\left(\frac{t}{t_{0}}\right)^{\frac{2}{3}},\qquad\rho_{b}(t)=\left(6\pi Gt^{2}\right)^{-1}\label{scale factor}\end{equation}
 where $t_{0}$ is some arbitrary initial time corresponding, say,
to the epoch of recombination, $G$ is the universal gravitational
constant, and $\rho_{b}(t)$ is the average, uniform, density frequently
referred to as the background density. These results can be obtained
directly from Eq.\ref{scaldis} by noting that if the density is uniform,
so that all matter is moving with the Hubble flow, the first two terms
in Eq.\ref{scaldis} vanish whereas the third term (times $A$) must
be equated to the gravitational field resulting from the uniformly
distributed mass contained within a sphere of radius $A\left|\mathbf{x}\right|$.
Then the third term of Eq.\ref{scaldis} is simply the contribution
arising by subtracting the field due to the background density from
the sphere.\cite{peebles2} Noting that $A^{3}\rho_{b}(t)=\rho_{b}(t_{0})$
forces the result. Alternatively, also for the case of uniform density,
taking the divergence of each side of Eq. \ref{scaldis} and asserting
the Poisson equation forces the same result. Thus the Friedman scaling
is consistent with the coupling of a uniform Hubble flow with Newtonian
dynamics.\cite{peebles2}

For computational purposes it is useful to obtain autonomous equations
of motion which do not depend explicitly on the time. This can be
effectively accomplished \cite{Rouet1,Rouet2} by transforming the
time coordinate according to

\begin{equation}
dt=B(t)d\tau,\qquad B(t)=\frac{t}{t_{0}}\label{timescal}\end{equation}
 yielding the autonomous equations

\begin{equation}
\frac{d^{2}\mathbf{x}}{d\tau^{2}}+\frac{1}{3t_{0}}\frac{d\mathbf{x}}{d\tau}-\frac{2}{9t_{0}^{2}}\mathbf{x=E}_{g}(\mathbf{x}).\label{auto}\end{equation}
For the Newtonian dynamics considered here it is necessary to confine
our attention to a bounded region of space, $\Omega$. It is customary
to choose a cube for $\Omega$ and assume periodic boundary conditions
. Thus our equations correspond to a dissipative dynamical system
in the comoving frame with friction constant $1/3t_{0}$ and with
forces arising from fluctuations in the local density with respect
to a uniform background.

For the special case of a stratified mass distribution the local density
at time $t$ is given by

\begin{equation}
\rho(x,t)=\sum m_{j}(t)\delta(x-x_{j})\label{den}\end{equation}
 where $m_{j}(t)$ is the mass per unit area of the $j^{th}$ sheet
and, from symmetry, the gravitational field only has a component in
the $x$ direction. Then, for the special case of equal masses $m_{j}(t)=m(t)$,
the correct form of the gravitational field occurring on the right
hand side of Eq \ref{auto} at the location of particle $i$ is then

\begin{equation}
E_{g}(x_{i})=2\pi m(t_{0})G[N_{R,i}-N_{L,i}]\label{field}\end{equation}
 since we already implicitly accounted for the fact that $m_{j}(t)=m(t_{0})/A^{2}$
in Eq.\ref{auto}. The equations are further simplified by establishing
the connection between $m(t_{0})$ and the background density at the
initial time, $\rho_{b}(t_{0}).$ Let us assume that we have $2N$
particles (sheets) confined within a slab with width $2L$, i.e. $-2L<x<2L$.
Then

\begin{equation}
\rho_{b}(t_{o})=\left(6\pi Gt_{o}^{2}\right)^{-1}=\left(\frac{N}{L}\right)m(t_{0}),\end{equation}
 we may express the field by

\begin{equation}
E_{g}(x_{i})=2\pi m(t_{0})G[N_{R,i}-N_{L,i}]=\frac{2}{3t_{0}^{2}}\left(\frac{L}{N}\right)[N_{R,i}-N_{L,i}],\end{equation}
 and the equations of motion for a particle in the system now read

\begin{equation}
\frac{d^{2}x_{i}}{d\tau^{2}}+\frac{1}{3t_{0}}\frac{dx_{i}}{d\tau}-\frac{2}{9t_{0}^{2}}x_{i}\mathbf{=}\frac{2}{3t_{0}^{2}}\left(\frac{L}{N}\right)[N_{R,i}-N_{L,i}].\end{equation}
 It is convenient to refer to Jeans theory for the final choice of
units of time and length \cite{b&t}\begin{eqnarray}
T_{j}=\omega_{j}^{-1}=(4\pi G\rho)^{-1/2}=\sqrt{\frac{3}{2}}t_{o}\nonumber \\
\quad\lambda_{j}=\lambda_{j}=v_{T}/\omega_{j}\sqrt{3\sigma_{v}^{2}/\omega_{j}^{2}}=\frac{3}{\sqrt{2}}\sigma_{v}t_{o},\label{jeans}\end{eqnarray}
 where $\sigma_{v}^{2}$ is the variance of the velocity at the initial
time, $v_{T}=\sigma_{v}=a/\sqrt{3}$ is the thermal velocity, and
$a$ is the maximal absolute velocity value given to a particle when
the initial distribution of velocities is uniform on $[-a,$$a$].
Then, with the further requirement that $L=n\lambda_{j},\quad0<n<N,$
in these units our equations are

\begin{equation}
\frac{d^{2}x_{i}}{d\tau^{2}}+\frac{1}{\sqrt{6}}\frac{dx_{i}}{d\tau}-\frac{1}{3}x_{i}\mathbf{=}\left(\frac{n}{N}\right)[N_{R,i}-N_{L,i}].\end{equation}

There is still a final issue that we have to address. If we assume
for the moment that the sheets are uniformly distributed and moving
with the Hubble flow, then the first two terms above are zero. Unfortunately,
if we take the average over each remaining term, we find that they
differ by a factor of three. This seeming discrepancy arises because
we started with spherical symmetry about an arbitrary point for the
Hubble flow and are now imposing axial symmetry. If we imagine that
we are in a local region with a stratified geometry, the symmetry
is different and this has to be reflected in the background term.
This apparent discrepancy can be rectified by multiplying the term
in $x_{i}$ by three. Our final equations of evolution are then

\begin{equation}
\frac{d^{2}x_{i}}{d\tau^{2}}+\frac{1}{\sqrt{6}}\frac{dx_{i}}{d\tau}-x_{i}\mathbf{=}\left(\frac{n}{N}\right)[N_{R,i}-N_{L,i}].\label{final}\end{equation}
 The description is completed by assuming that the system satisfies
periodic boundary conditions on the interval $2L$, i.e. when a particle
leaves the primitive cell defined by $-2L<x<2L$ on the right, it
re-enters at the left hand boundary with the identical velocity. Note
that, in computing the field, we do not attempt to include contributions
from the images of $x_{i}$ outside of the primitive cell.

Finally, we mention that the RF model is obtained from the reverse
sequence where one first restricts the geometry to 1+1 dimensions
and then introduces the transformation to the comoving frame. In this
approach it is not necessary to make the adjustment in the coefficient
of $x_{i}$ as we did here. This is quickly seen by noting that the
divergence of $\mathbf{x}$ is three times greater than the divergence
of $x\hat{x}$ which one would obtain by directly starting with the
one dimensional model. However, in the RF model, the coefficient of
the first derivative term (the friction constant) is $1/\sqrt{2}$
instead of $1/\sqrt{6}$. This simply illustrates that, since there
is no curvature in a 1+1 dimensional universe, there is a degree of
arbitrariness in choosing the final model. It cannot be obtained solely
from General Relativity. For a discussion of this point see Mann et.
al. . \cite{Mann1} A Hamiltonian version can also be considered by
setting the friction constant to zero. In their earlier work, using
the linearized Vlasov-Poisson equations, Rouet and Feix carried out
a stability analysis of the model without friction. They determined
that the system followed the expected behavior, i.e. when the system
size is greater than the Jean's length, instability occurs and clustering
becomes possible.\cite{Rouet1,Rouet2} We mention that, when the friction
term is not present, both the Q and RF models are identical so, with
the assumption that the friction term won't have a large influence
on short time, linear, stability, the analysis of the Hamiltonian
version applies equally to each version.

The Vlasov-Poisson limit for the system is obtained by letting $N\rightarrow\infty$
and $m\rightarrow0$ while constraining the density and energy (at
a given time). Then, in the comoving frame, the system is represented
by a fluid in the $\mu$ space. Let $f(x,v;t)$ represent the normalized
distribution of mass in the $(x,v)$ phase plane at time $t$. From
mass conservation, $f$ satisfies the continuity equation

\begin{equation}
\frac{\partial f}{\partial t}+v\frac{\partial f}{\partial x}+\frac{\partial af}{\partial v}=0,\quad a=-\gamma v-\frac{\partial\phi_{T}(x)}{\partial x}\label{cont}\end{equation}

\noindent which can also be derived using the identical scaling as
aboveS. In Eq. (\ref{cont}), $a(x,v)$ is the local acceleration
and $\phi_{T}$ is the potential function induced by the total density,
including the effective negative background density, $-\rho_{b}$.
Note that both $a(x,v)$ and $\phi_{T}$ are linear functionals of
$f(x,v;t)$. Coupling Eq. (\ref{cont}) with the Poisson equation
yields the complete Vlasov-Poisson system governing the evolution
of $f(x,v;t)$. Depending on the final choice of the friction constant,
$\gamma$, the continuum limit of either the Quintic or RF model can
be represented by Eq.\ref{cont}. Note that an exact solution is

\begin{equation}
f(x,v,t)=\frac{|\rho_{b}|}{\sqrt{2\pi\sigma^{2}}}\exp{-\frac{v^{2}}{2\sigma^{2}}},\quad\sigma(t)=\sigma_{0}\exp{-\gamma t}\label{GaussVlas}\end{equation}

\noindent Using dynamical simulation, we will see that it is extremely
unstable when the system size exceeds the Jeans' length at the initial
time.

Useful information can be had without constructing an explicit solution.
For example, we quickly find that the system energy decreases at a
rate proportional to the kinetic energy. The Tsallis entropy is defined
by

\begin{equation}
S_{q}=\frac{1-\int{f^{q}dxdv}}{q-1},\quad\int\int{\ fdxdv}=1\label{Tsallis}\end{equation}

\noindent In the limit $q\rightarrow1$, $S_{q}$ reduces to the usual
Gibbs entropy, $S_{1}=-\int{\int{f\ln{f}dx}dv}$. By asserting the
Vlasov-Poisson evolution, we find that the Bolztmann-Gibbs entropy,
$S_{1}$, decreases at the constant rate $-2\gamma$ while, for $q>1$,
$S_{q}$ decreases exponentially in time. This tells us that the mass
is being concentrated in regions of decreasing area of the phase plane,
suggesting the development of structure. These properties are immediately
evident for the trivial solution given above. By imposing a Euclidean
metric in the phase plane, we can also investigate local properties
such as the directions of maximum stretching and compression, as well
as the local vorticity. We easily find that the rate of separation
between two nearby points is a maximum in the direction given by

\begin{equation}
\tan(2\theta)=(1+\rho+\rho_{b})/\gamma,\label{stretching}\end{equation}

\noindent where $\theta$ defines the angle with the coordinate axis
in the $\mu$ space. Thus, in regions of low density, we expect to
see lines of mass being stretched with constant positive slope in
the phase plane. We will see below that this prediction is accurately
born out by simulations.

\section{Simulations}

\label{sim} An attraction of these one-dimensional gravitational
systems is their ease of simulation. In both the autonomous (purely
Newtonian without expansion) and RF models it is possible to integrate
the motion of the individual particles between crossings analytically.
Then the temporal evolution of the system can be obtained by following
the successive crossings of the individual, adjacent, particle trajectories.
This is true as well for the Q model. If we let $y_{i}=x_{i+1}-x_{i}$,
where we have ordered the particle labels from left to right, then
we find that the differential equation for each $y_{i}$ is the same,
namely

\begin{equation}
\frac{d^{2}y_{i}}{d\tau^{2}}+\frac{1}{\sqrt{6}}\frac{dy_{i}}{d\tau}-y_{i}\mathbf{=-}\frac{2n}{N}.\label{disp}\end{equation}

The general solution of the homogenious version of E. \ref{disp}
is a sum of exponentials. By including the particular solution of
the inhomogenius equation (simply a constant) we obtain a fifth order
algebraic equation in $u=\exp(\tau/\sqrt{6})$. Hence the name Q,
or Quintic, model. These can be solved numerically in terms of the
initial conditions by analytically bounding the roots and employing
the Newton-Raphson method.\cite{recipes_in_C} (Note that for the
RF model a cubic equation is obtained.) A sophisticated, event driven,
algorithm was designed to execute the simulations. Two important features
of the algorithm are that it only updates the positions and velocities
of a pair of particles when they actually cross, and it maintains
the correct ordering of each particle's position on the line. Using
this algorithm we were able to carry out runs for significant cosmological
time with large numbers of particles.

Typical numerical simulations were carried out for systems with up
to $N=2^{18}$ particles. Initial conditions were chosen by equally
spacing the particles on the line and randomly choosing their velocities
from a uniform distribution within a fixed interval. For historical
reasons we call this a waterbag. Other initial conditions, such as
Normally distributed velocities, as well as a Brownian motion representation,
which more closely imitates a cosmological setting, were also investigated.
The simulations were carried forward for approximately fifteen dimensionless
time units.

\begin{figure*}[!ht]
 \centerline{\includegraphics[width=1\textwidth]{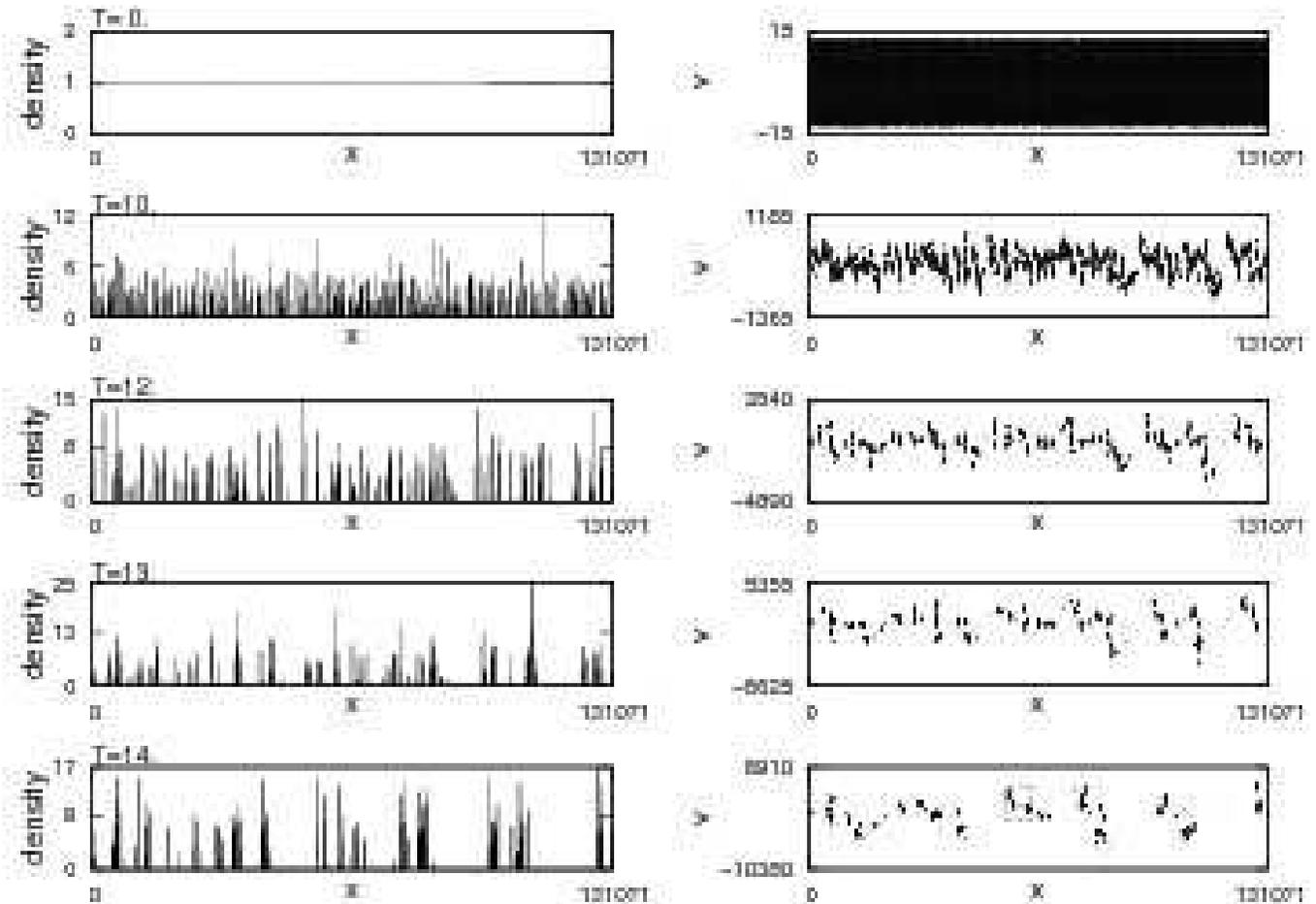}}

\caption{\label{sim1} Evolution in configuration and $\mu$ space for the
quintic model with $2^{17}$ particles from T=0 to T=14. The initial
distribution is a waterbag with velocities in the range (-12.5, 12.5)
and a size of about 16,000 Jeans' lengths.}
\end{figure*}

\begin{figure}[!ht]
 \includegraphics[width=0.48\textwidth]{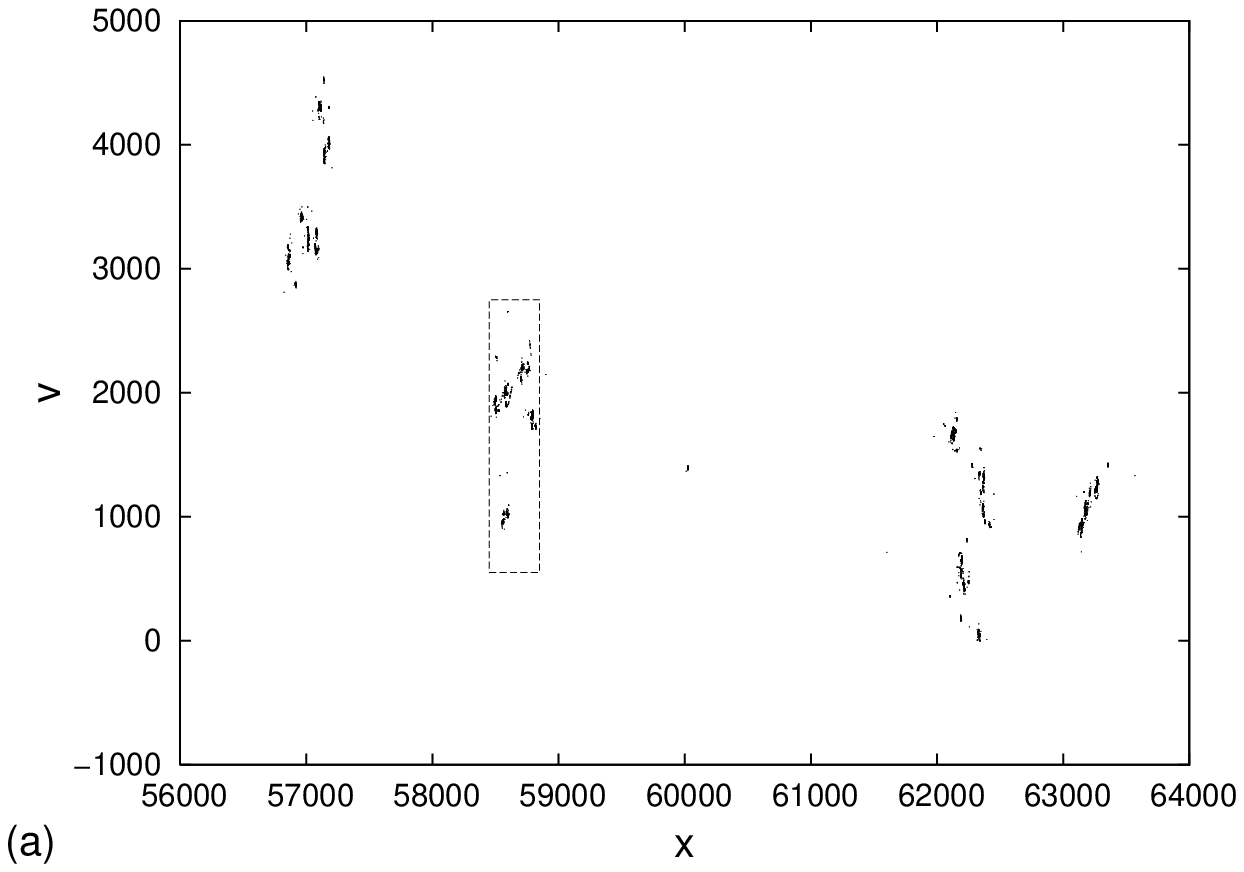}
\includegraphics[width=0.48\textwidth]{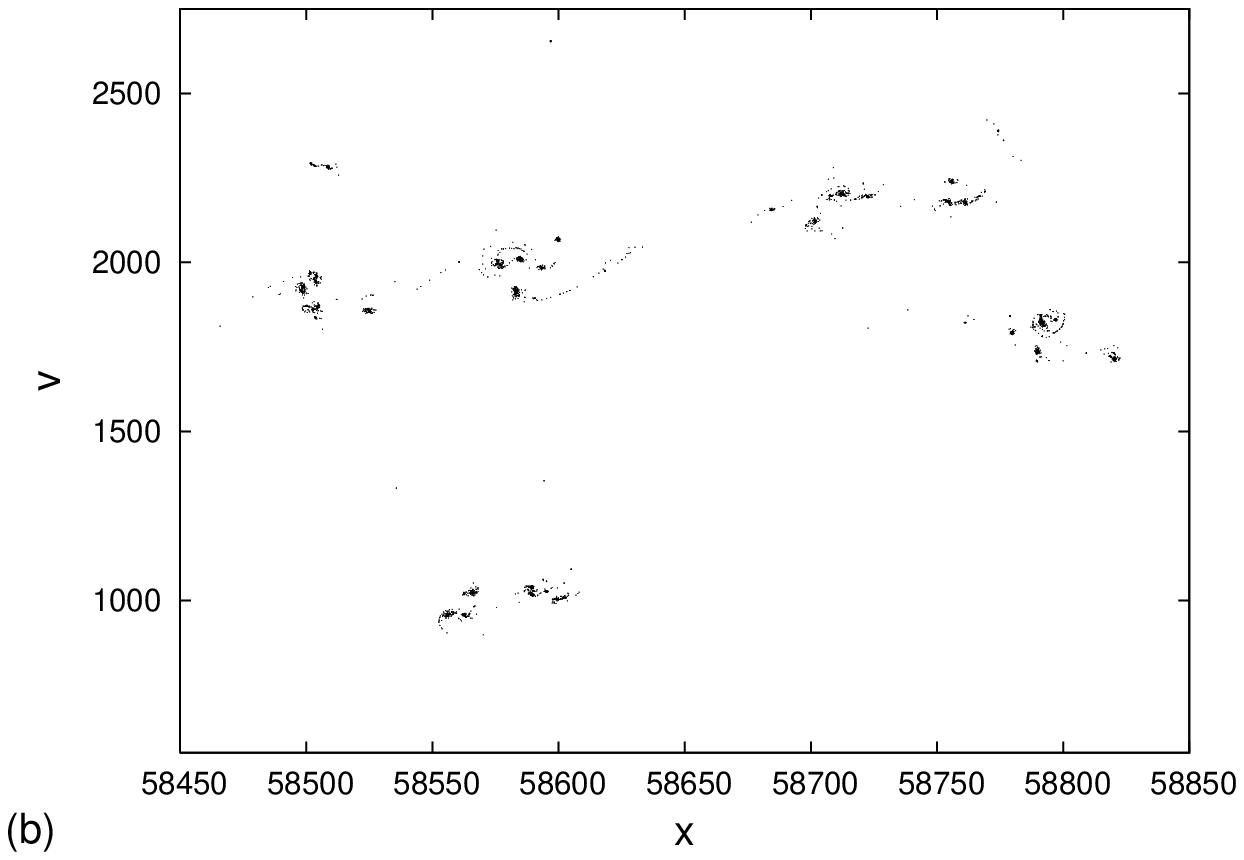}

\caption{\label{zoom1} Consecutive expansions (zooms) on successive small
squares selected in the $\mu$ space panels at T=14 for the quintic
model. They have the appearance of a random fractal which suggests
self-similarity.}
\end{figure}

In Fig. \ref{sim1} we present a visualization of a typical run with
$2^{17}$particles. The system consists of the Q model with an initial
waterbag distribution in the $\mu$ space. Initially the velocity
spread is (-12.5, 12.5) in the dimensionless units employed here and
the system contains approximately 16,000 Jeans' lengths. In the left
column we present a histogram of the particles positions at increasing
time frames, while on the right we display the corresponding particle
locations in $\mu$ (position, velocity) space. It is clear from the
panels that hierarchical clustering is occurring, i.e. small clusters
are joining together to form larger ones, so the clustering mechanism
is {}``bottoms up''\cite{peebles2}. The first clusters are roughly
the size of a Jean's length and seem to appear at about $\tau=6$
and there are many while by $\tau=14$ there are on the order of 30
clusters. In the $\mu$ space we observe that between the clusters
matter is distributed along linear paths. As time progresses the size
of the linear segments increases. The behavior of these under-dense
regions is governed by the stretching in $\mu$ space predicted by
Vlasov theory explained above (see before Eq.(\ref{stretching})).
The slopes of the segments are the same and in quantitative agreement
with Eq. (\ref{stretching}). Qualitatively similar histories are
obtained for the RF model and the model without friction (see below),
as well as for the different boundary conditions mentioned above.
However, there are some subtle differences. In Fig. \ref{zoom1} we
zoom in and show a sequence of magnified inserts from the $\mu$space
at time T=14. The hierarchical structure observed in these models
suggests the existence of a fractal structure, but careful analysis
is required to determine if this is correct. Simulations have also
been performed where the system size is less than the Jean's length.
The results support the standard stability analysis in that hierarchical
clustering is not observed for these initial conditions.

\begin{figure}[!ht]
 \centerline{\includegraphics[width=0.48\textwidth]{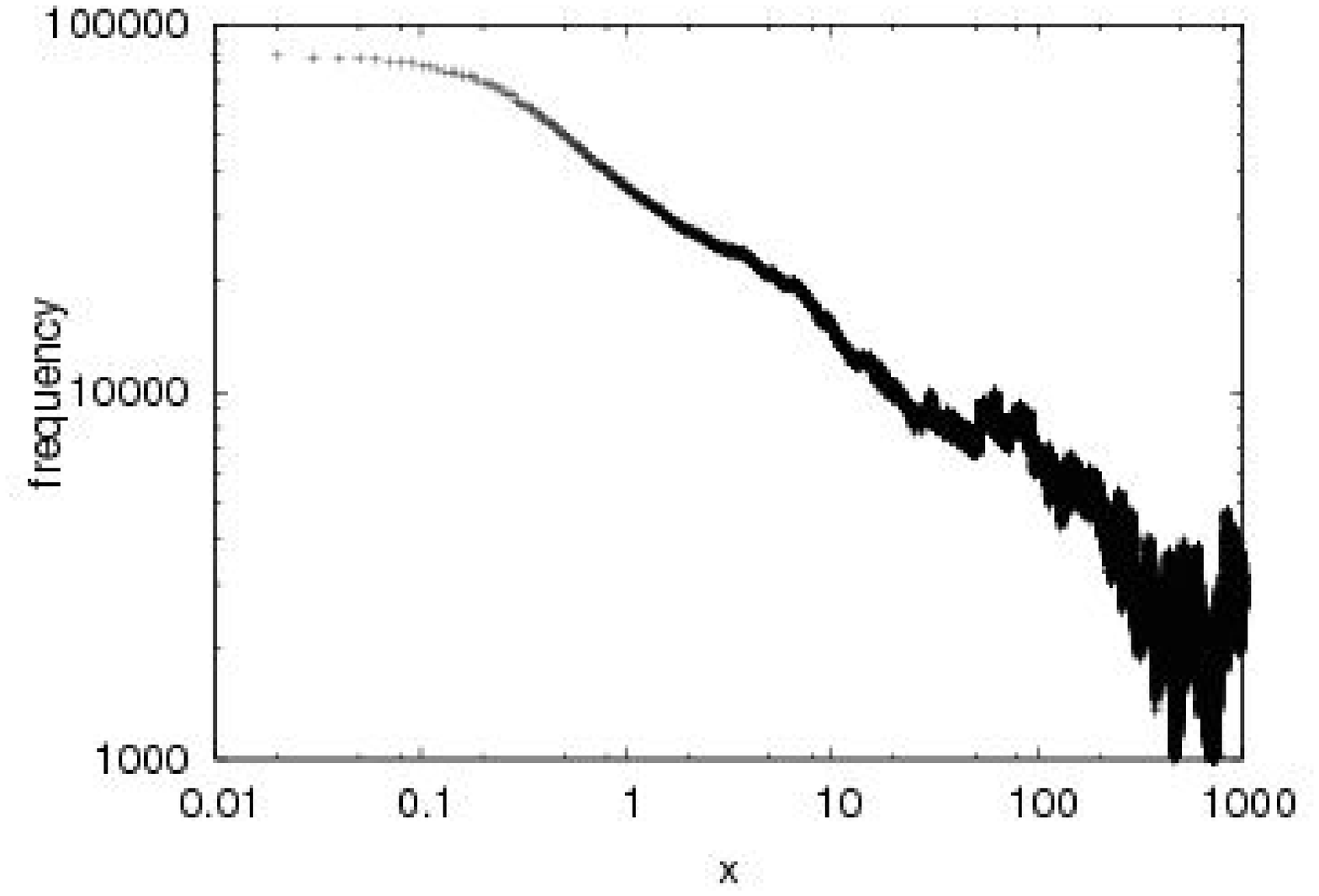}}

\caption{\label{cor_fun}Figure 3. The correlation function at T=14 which
shows a scaling region from about 0.3 to about 30, a range of about
2 decades with a scaling exponent $\gamma=.42$.}
\end{figure}

Historically, power law behavior in the density-density correlation
funtion has been taken as the most important signature of self similar
behavior of the distribution of galaxy positions.\cite{cosmo-rev}
In Figure \ref{cor_fun} we provide a log-log plot of the correlation
function $C(r)$ at T=14 defined by

\begin{equation*}
C(r)=<\delta\rho(y+r)\delta\rho(y)>\sim\frac{1}{\Delta}\int_{r}^{r+\Delta}\left[\sum_{i,j,i\neq j}\delta(r'-(y_{i}-y_{j}))-\frac{N\left(N-1\right)}{L^{2}}\right]dr'
\end{equation*}

\label{cor_eq} where particles $i$ and $j$ are such that $L/4<x_{i,j}<3L/4$
to avoid boundary effects and $\Delta$ is the bin size. Note the
existence of a scaling region from about $0.1<\ln(l)<30$, a range
of about 2.5 decades in $l$. Note also the noise present at larger
scales. Since the fluctuations are vanishingly small at scales on
the order of the system size, this is most likely attributable to
the presence of shot noise. This could be reduced by taking an ensemble
average over many runs. By computing the slope, say $\gamma$ , of
the log-log plot in Fig. \ref{cor_fun} we are able to obtain an estimate
of the correlation dimension $D_{2}=1-\gamma$ for a one dimensional
system. We find that $\gamma=.42$ for a scaling region of about two
decades in $l$ . This suggests that the correlation dimension is
approximately 0.6, which is in agreement within the standard numerical
error with the multifractal analysis described table \ref{tab1} below in some detail.

\section{Fractal Measures}

\label{frac} It is natural to assume that the apparently self similar
structure which develops in the phase plane (see Figs.\ref{sim1},\ref{zoom1})
as time evolves has fractal geometry, but we will see that things
aren't so simple. In their earlier study of the RF model, Rouet and
Feix found a box counting dimension for the particle positions of
about 0.6 for an initial waterbag distribution (uniform on a rectangle
in the phase plane-see above) and a fractal dimension of about 0.8
in the configuration space (i.e. of the projection of the set of points
in $\mu$ space on the position axis). \cite{Rouet2}. As far as we
know, Bailin and Shaeffer were the first to suggest that the distribution
of galaxy positions are consistent with a bifractal geometry \cite{bs}
. Their idea was that the geometry of the galaxy distribution was
different in the clusters and voids and, as a first approximation,
this could be represented as a superposition of two independent fractals.
Of course, their analysis was restricted solely to galaxy positions.
Since the structures which evolve are strongly inhomogeneous, to gain
further insight we have carried out a multi-fractal analysis \cite{Hal}
in both the phase plane and the position coordinate.

The multifractal formalism shares a number of features with thermodynamics.
To implement it we partitioned each space configuration space and
$\mu$ space, into cells of length $l$. At each time of observation
in the simulation, a measure $\mu_{i}=N_{i}(t)/N$ was assigned to
cell $i$, where $N_{i}(t)$ is the population of cell $i$ at time
$t$ and $N$ is the total number of particles in the simulation.
The generalized dimension of order $q$ is defined by \cite{Hal}

\begin{equation}
D_{q}=\frac{1}{q-1}\lim_{l\rightarrow0}\frac{\ln{C_{q}}}{\ln{l}},\quad C_{q}=\Sigma\mu_{i}^{q}.\label{Dq}\end{equation}
 where $C_{q}(l)$ is the effective partition function \cite{Ott},
$D_{0}$ is the box counting dimension, $D_{1}$, obtained by taking
the limit $q\rightarrow1$, is the information dimension, and $D_{2}$
is the correlation dimension \cite{Hal}\cite{Ott}. As $q$ increases
above $0$, the $D_{q}$ provide information on the geometry of cells
with higher population.

In practice, it is not possible to take the limit $l\rightarrow0$
with a finite sample. Instead, one looks for a scaling relation over
a substantial range of $\ln l$ with the expectation that a linear
relation between $\ln{C_{q}}$ and $\ln l$ occurs, suggesting power
law dependence of ${C_{q}}$ on $l$. Then, in the most favorable
case, the slope of the linear region should provide the correct power
and, after dividing by $q-1$, the generalized dimension $D_{q}$.
As a rule, or guide, if scaling can be found either from observation
or computation over three decades of $l$, then we typically infer
that there is good evidence of fractal structure.\cite{Mac} Also
of interest is the global scaling index $\tau_{q}$, where ${C_{q}\sim}l^{\tau_{q}}$
for small $l$ . It can be shown that $\tau_{q}$ and $D_{q}$ are
related to each other through a Legendre transformation by $D_{q}(1-q)=\tau_{q}$
for $q\neq1$ \cite{Ott}. Here we present the results of our fractal
analysis of the particle positions on the line (position only) and
the plane (position, velocity or $\mu$ space).

If it exists, a scaling range of $l$ is defined as the interval on
which plots of $\ln C_{q}$ versus $\ln l$ are linear. Of course,
for the special case of $q=1$ we plot $-\Sigma\mu_{i}ln(\mu_{i})$
vs. $ln(l)$ to obtain the information dimension. \cite{Ott} If a
scaling range can be found, $D_{q}$ is obtained by taking the appropriate
derivative. It is well established by proof and example that, for
a normal, homogeneous, fractal, all of the generalized dimensions
are equal, while for an inhomogeneous fractal, e.g. the Henon attractor,
$D_{q+1}\leq{D_{q}}$ \cite{Hal}. In the limit of small $l$, the
partition function, $C_{q}(l)$, can also be decomposed into a sum
of contributions from regions of the inhomogeneous fractal sharing
the same point-wise dimension, $\alpha$,

\begin{equation}
C_{q}(l)=\int d\alpha l^{\alpha q}\rho(\alpha)\exp\left[-f\left(\alpha\right)\right]\end{equation}
 where $f\left(\alpha\right)$ is the fractal dimension of its support
\cite{Hal}\cite{Fed}\cite{Ott}. Then if, for a range of $q$, a
single region is dominant, we find a simple relation between the global
index, $\tau_{q}$, and $\alpha$,

\begin{equation}
\tau_{q}=\alpha q-f(\alpha).\end{equation}

Recall that a Euclidean metric is imposed on the $\mu$-space. Because,
in our units, $\omega_{j}=1,$ we divide it into cells such that $\Delta x=\Delta v$.
Then, as $L$ is large ($L\approx10,000$), so is the extent of the
partition in the velocity space. On this scale the initial distribution
of particules appears as a line so that the initial dimension is also
about unity. In fact, initially the velocity of the particles is a
perturbation. It is not large enough to allow particles to cross the
entire system in a unit of time. During the initial period the virial
ratio rapidly oscillates. After a while the granularity of the system
destroys the approximate symmetry of the initial $\mu$-space distribution.
Breaking the symmetry leads to the short time dissipative mixing that
results in the separation of the system into clusters. Although the
embedding dimension is different, the behavior of the distribution
of points in configuration space is similar. The initial dimension
is nearly one until clustering commences. At this time the dimensions
in $\mu$-space and configuration space separate.

As time progresses, however, for the initial conditions discussed
above, for $q>0$ typically two independent scaling regimes developed.
Of course this is in addition to the trivial scaling regions obtained
for very small $l$ , corresponding to isolated points, and to large
$l$ on the order of the system size, for which the matter distribution
looks smooth. The observed size of each scaling range depended on
both the elapsed time into the simulation and the value of $q$. While
the length of each scaling regime varied with both $q$ and time,
in some instances it was possible to find good scaling over up to
four decades in $l$!

\begin{figure}[!ht]
 \includegraphics[width=0.4\textwidth]{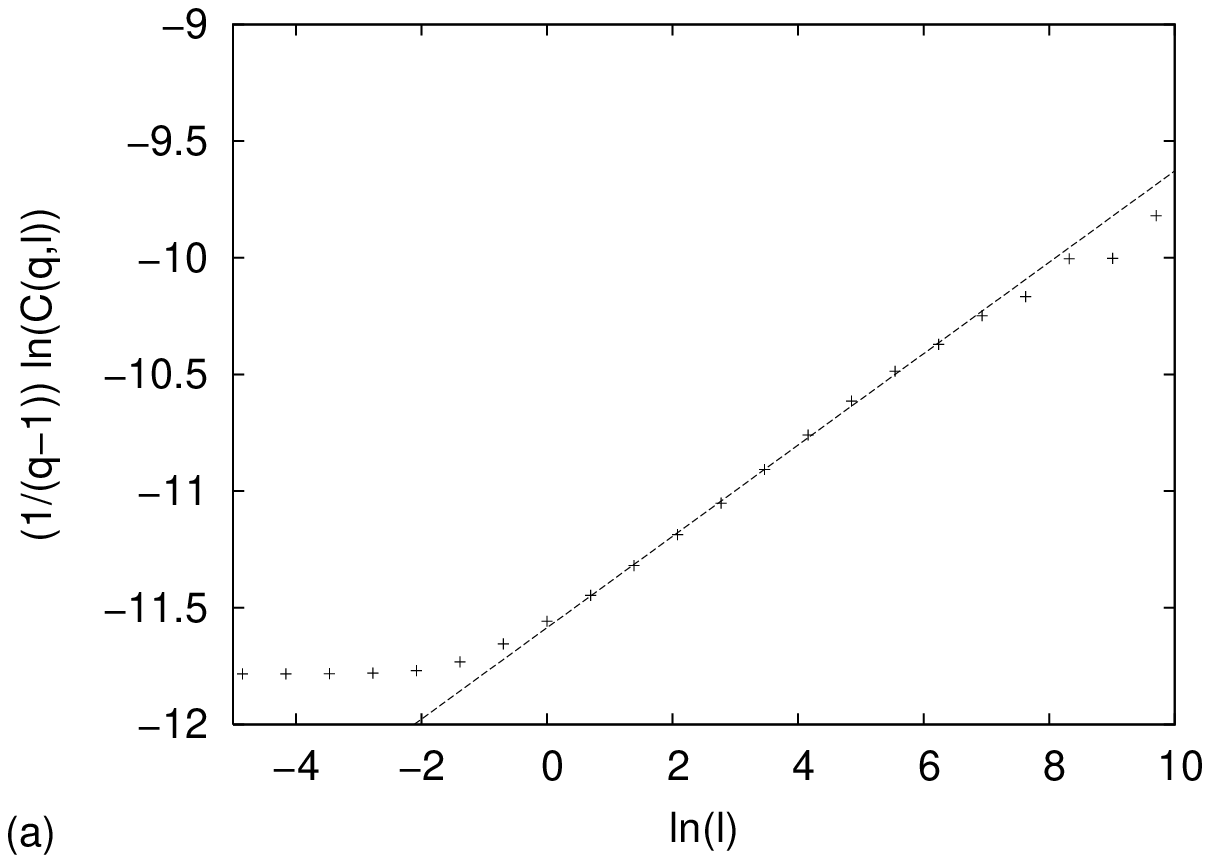}
\includegraphics[width=0.4\textwidth]{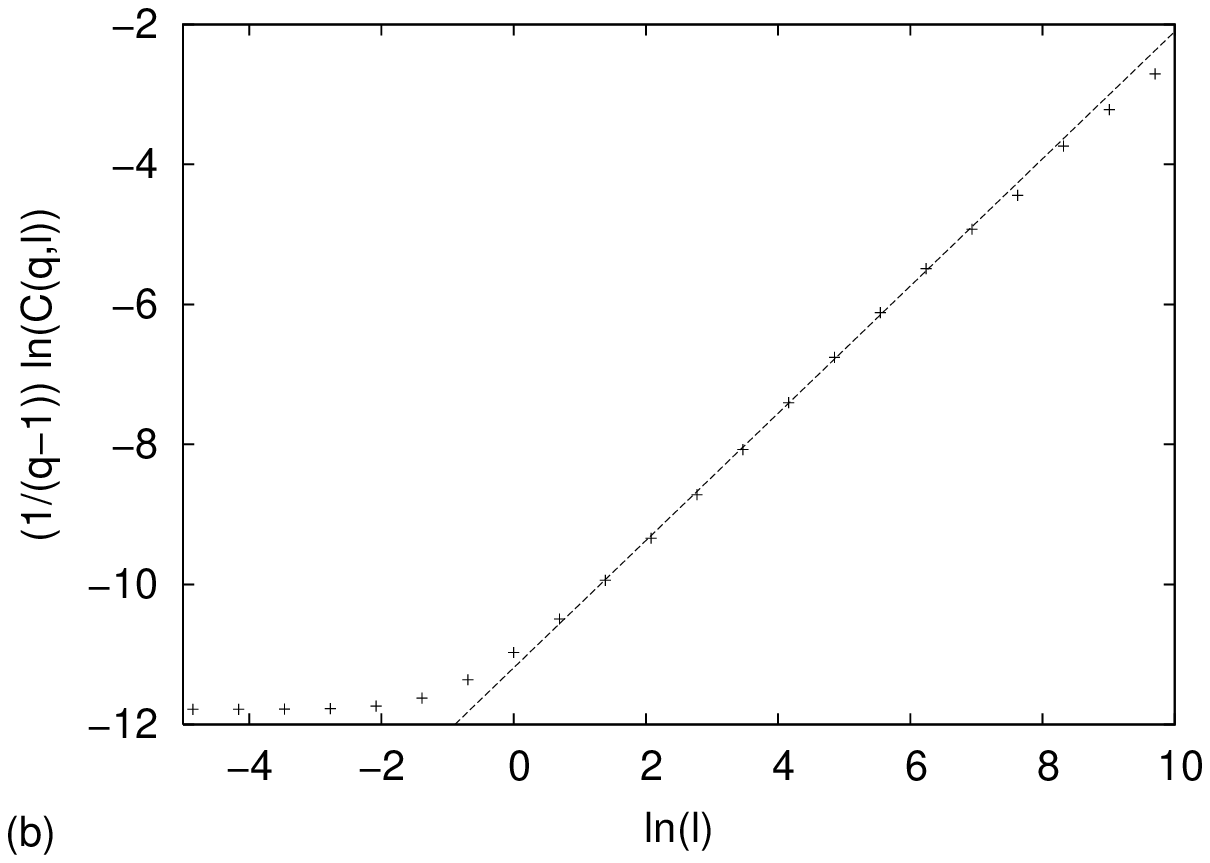}
\includegraphics[width=0.4\textwidth]{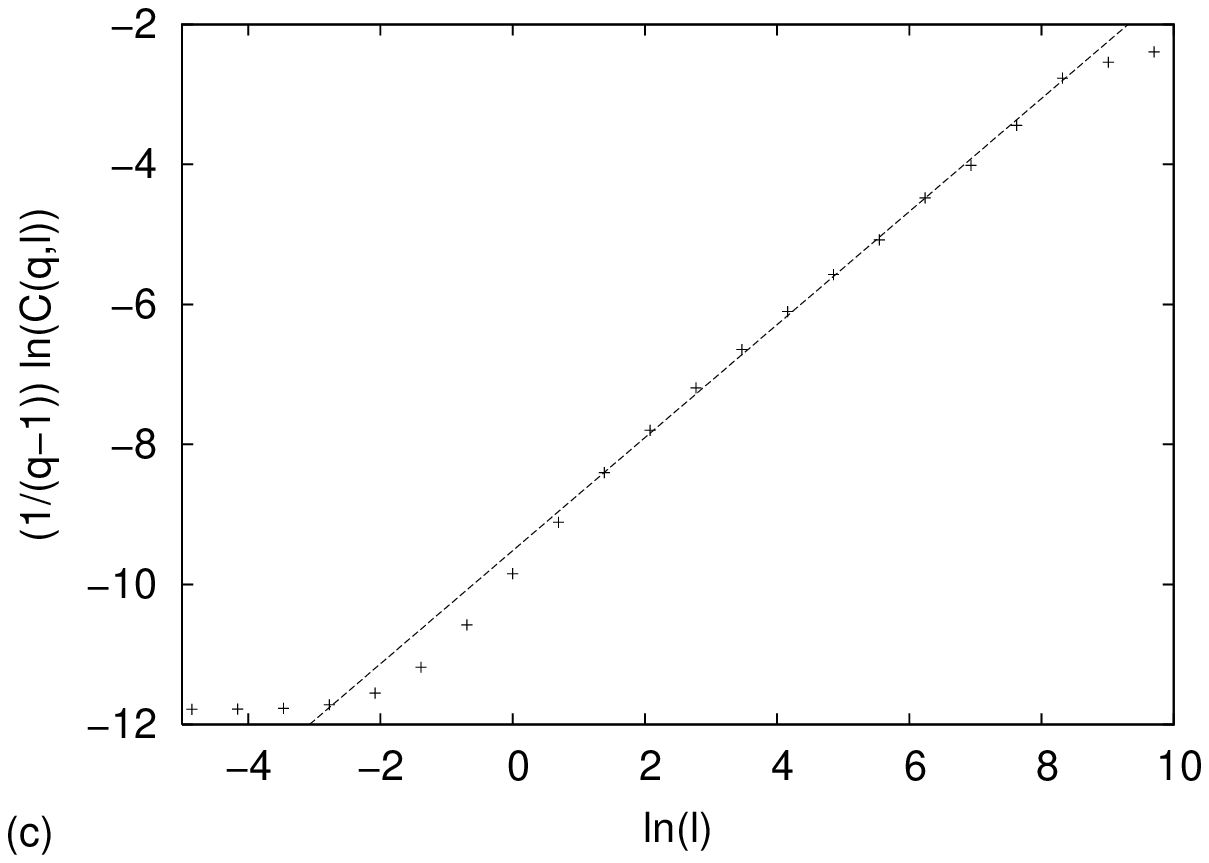}
\includegraphics[width=0.4\textwidth]{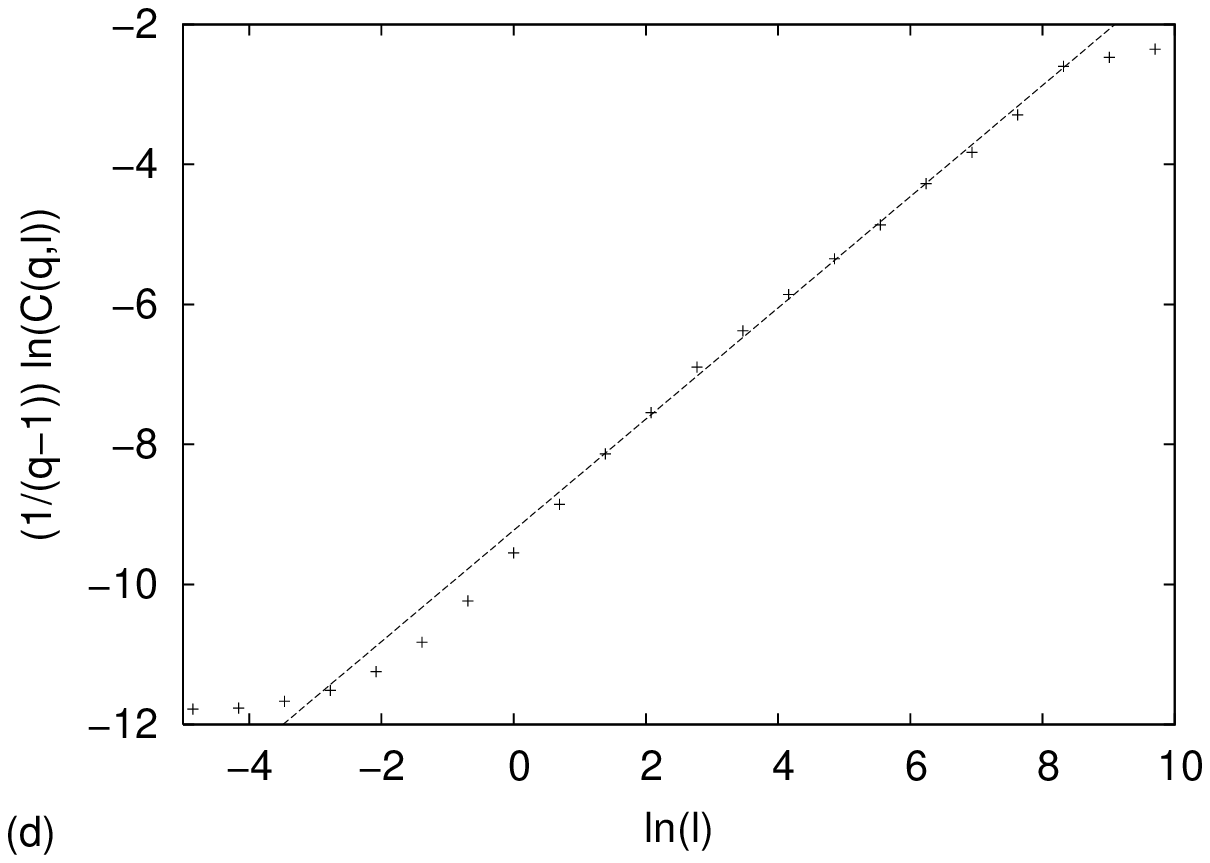}

\caption{\label{fig4}Scaling behavior in $\mu$-space. Plots of $\frac{1}{q-1}{\ln{C_{q}}}$
versus $\ln{l}$ are provided for four values of q: a) q=-5, b) q=0.
c) q=5, d) q= 10.}
\end{figure}

In Fig. \ref{fig4} we provide plots of $\frac{1}{q-1}{\ln{C_{q}}}$
versus $\ln{l}$ in $\mu$-space for four different values of $q$
covering most of the range we investigated ($-5<q<10$). To guarantee
that the fractal structure was fully developed, we chose $T=14$ for
the time of observation and the initial conditions are those given
above. For $q=-5$ and $q=0$ we clearly observe a single, large,
scaling range where $0.5<\ln{l}<8$ , corresponding to about three
decades in $l$. In the remaining panels (c, d), where we increase
$q$ to $5$ , and then $10$, we see a dramatic change. The large
scaling range has split into two smaller regions separated at about
$l=1.5$, and the slope of the region with larger $l$ has decreased
compared with the scaling region on its left ($-1.5<\ln{l}<1.5$).
The scaling range with larger $l$, $1.5<\ln{l}<8.5,$ corresponds
to just over three decades in $l$. Note that in panels c and d of
Fig. \ref{fig4}, the scaling range with larger $l$ is more robust.

\begin{figure}[!ht]
 \includegraphics[width=0.48\textwidth]{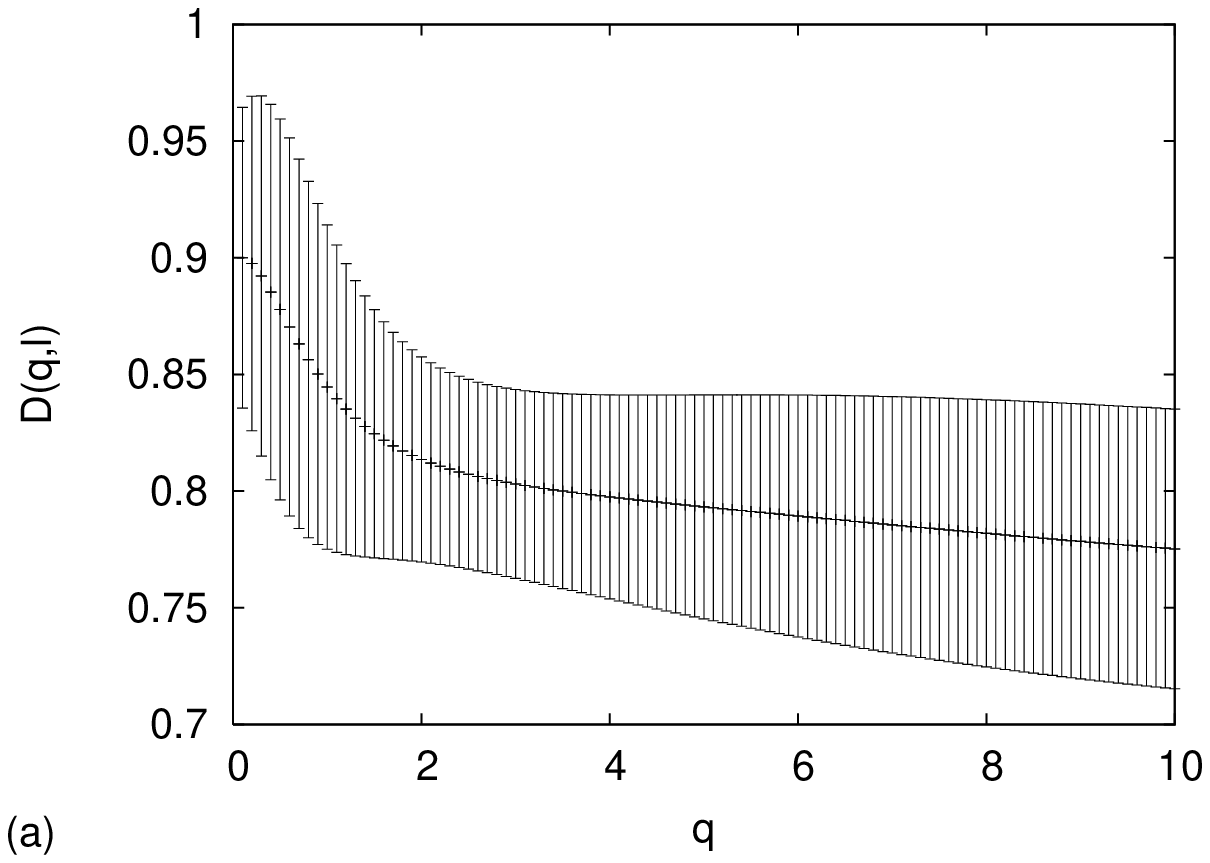}
\includegraphics[width=0.48\textwidth]{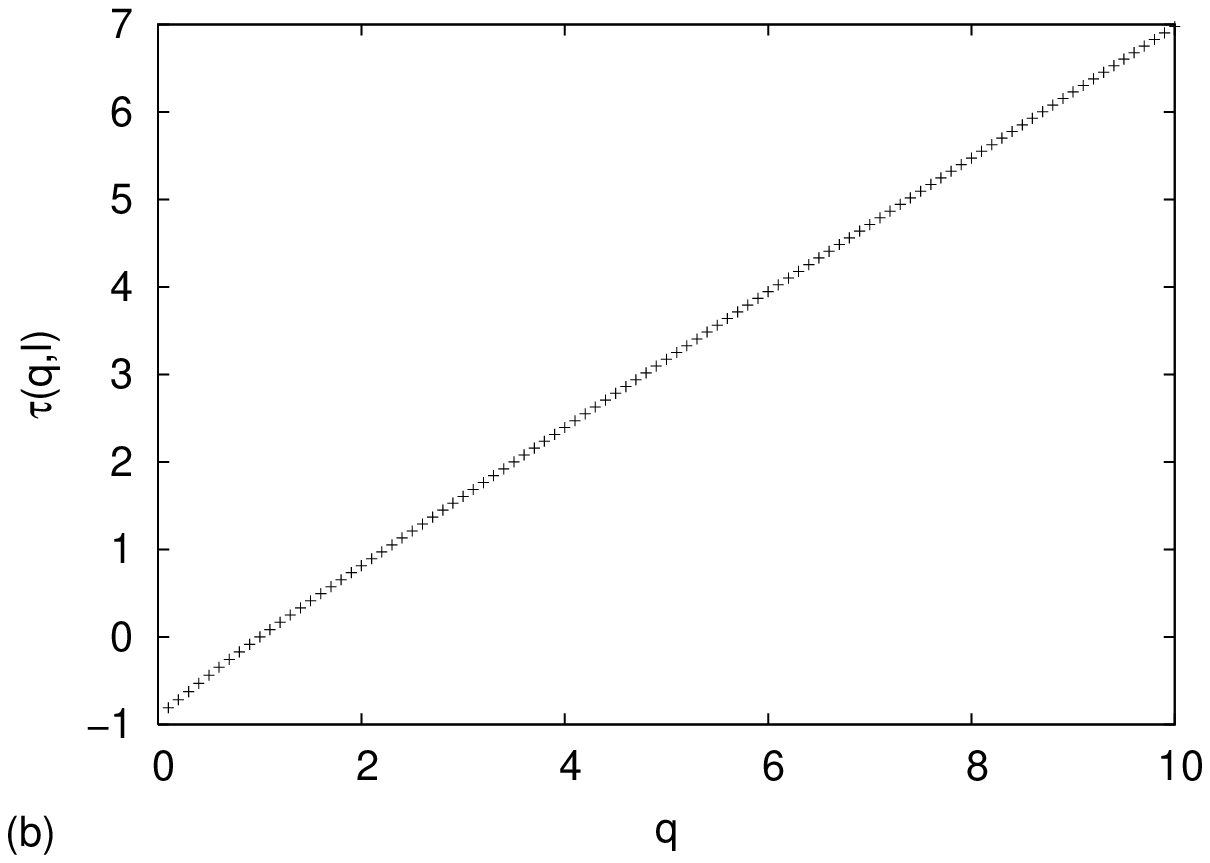}

\caption{\label{fig5} Generalized dimension $D_{q}$ vs. $q$ (panel a) and
global scaling index $\tau_{q}$vs. $q$ (panel b) in $\mu$-space
for the quintic model at T=14 with $q>0$.}
\end{figure}

\begin{figure}[!ht]
 \includegraphics[width=0.4\textwidth]{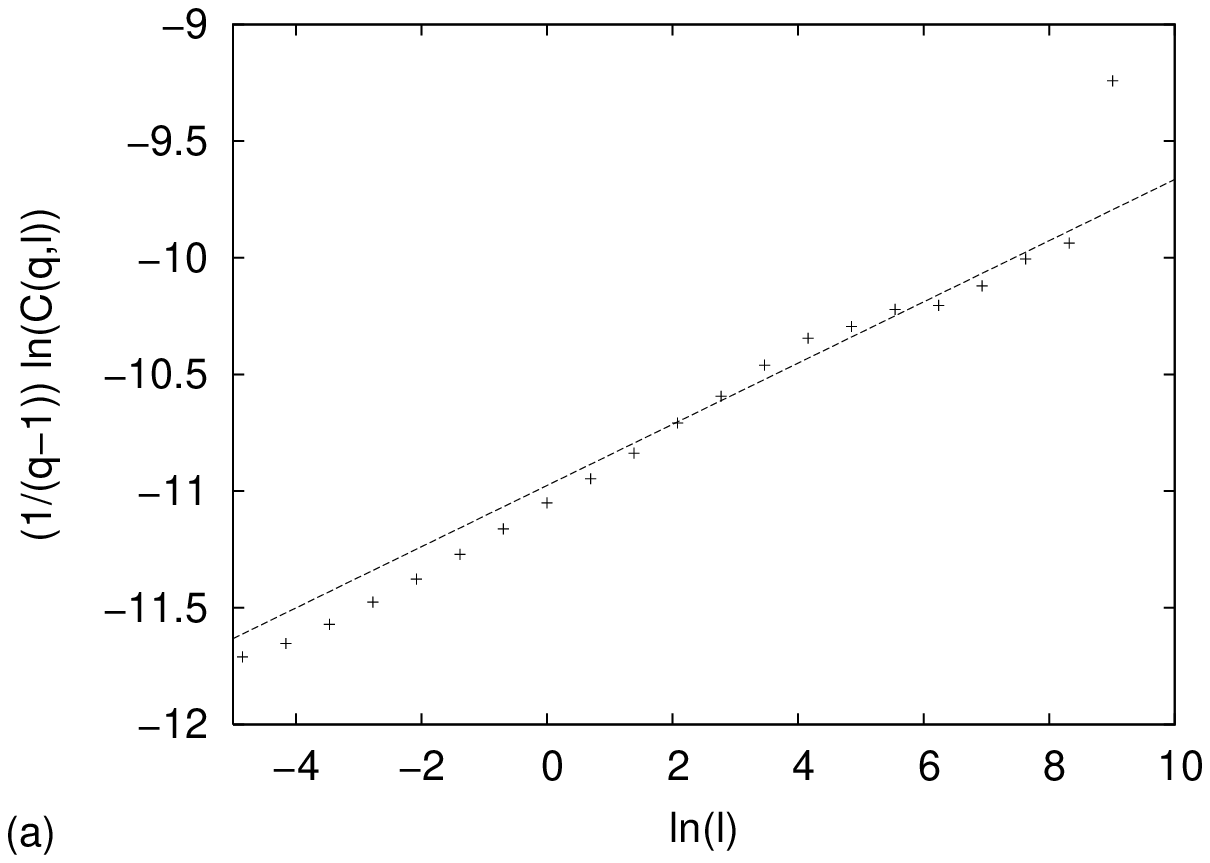}
\includegraphics[width=0.4\textwidth]{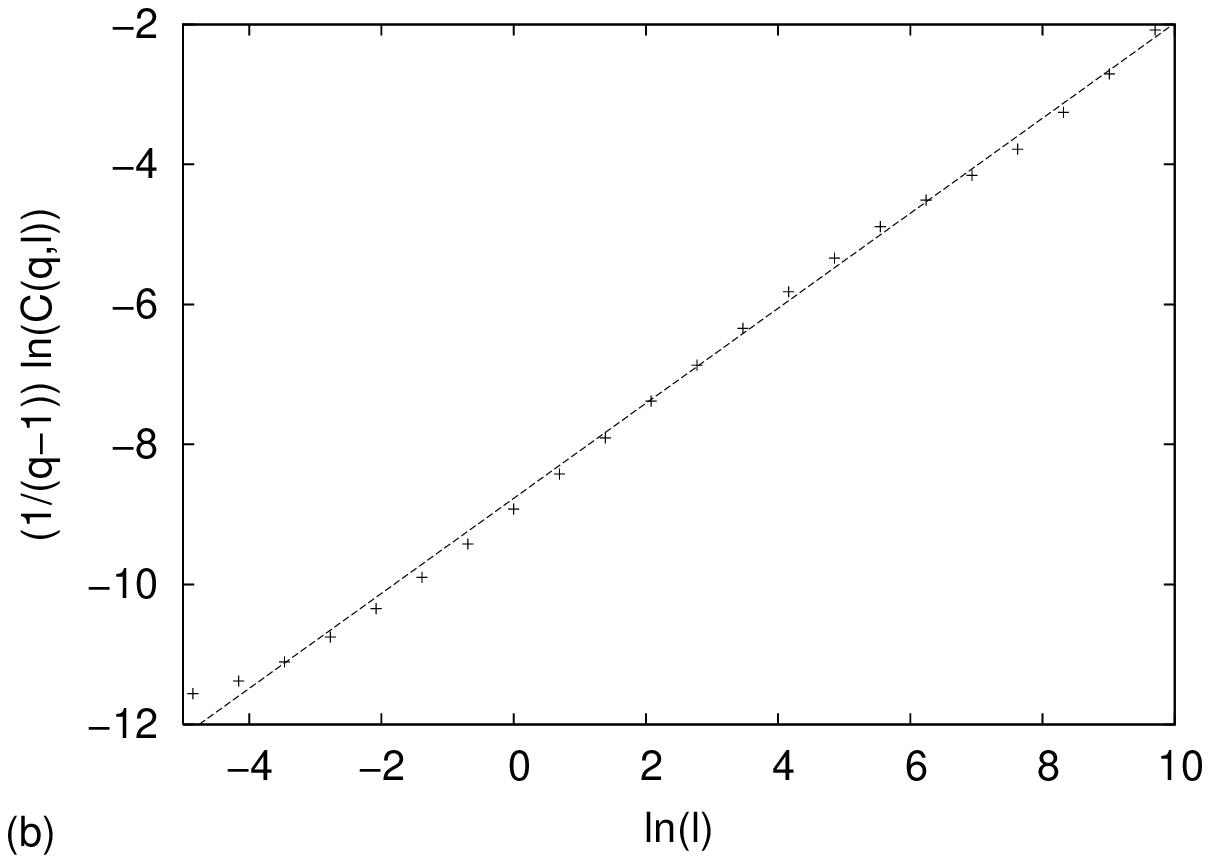}
\includegraphics[width=0.4\textwidth]{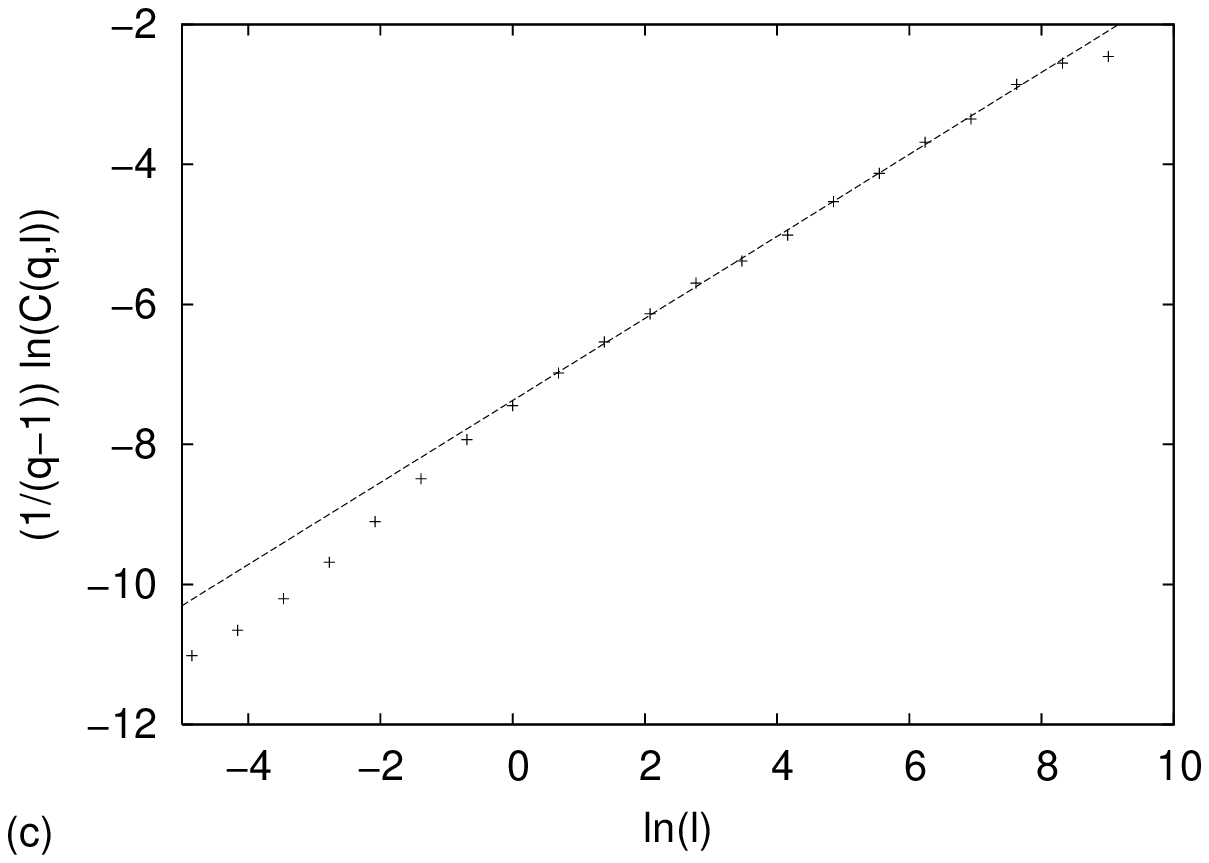}
\includegraphics[width=0.4\textwidth]{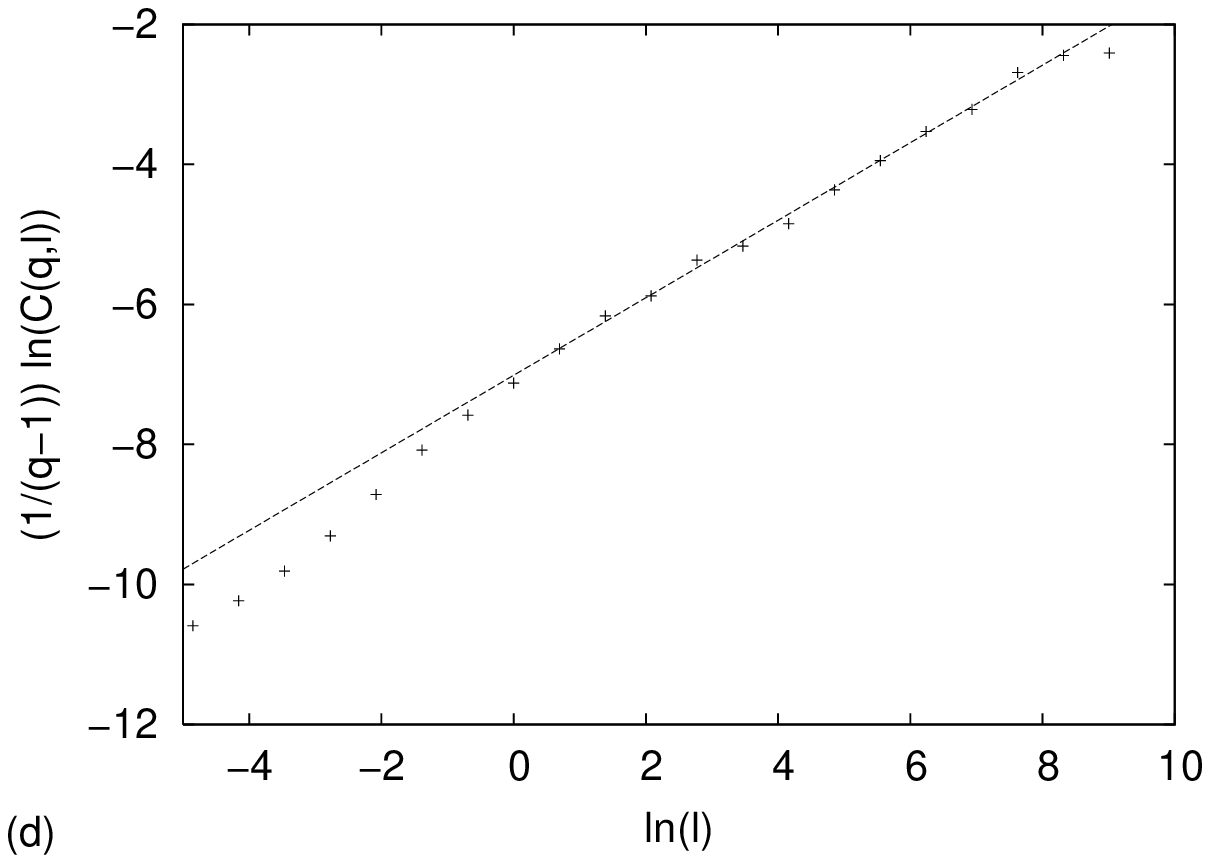}

\caption{\label{fig6} Plots of $1/(q-1)ln(C_{q})$ vs $ln(l)$ in configuration
($x$ ) space for the quintic model prepared as above at T=14 are
provided for four values of $q$: a) $q=-5$, b) $q=0$, c) $q=5$,
d) $q=10$.}
\end{figure}

\begin{figure}[!ht]
 \includegraphics[width=0.48\textwidth]{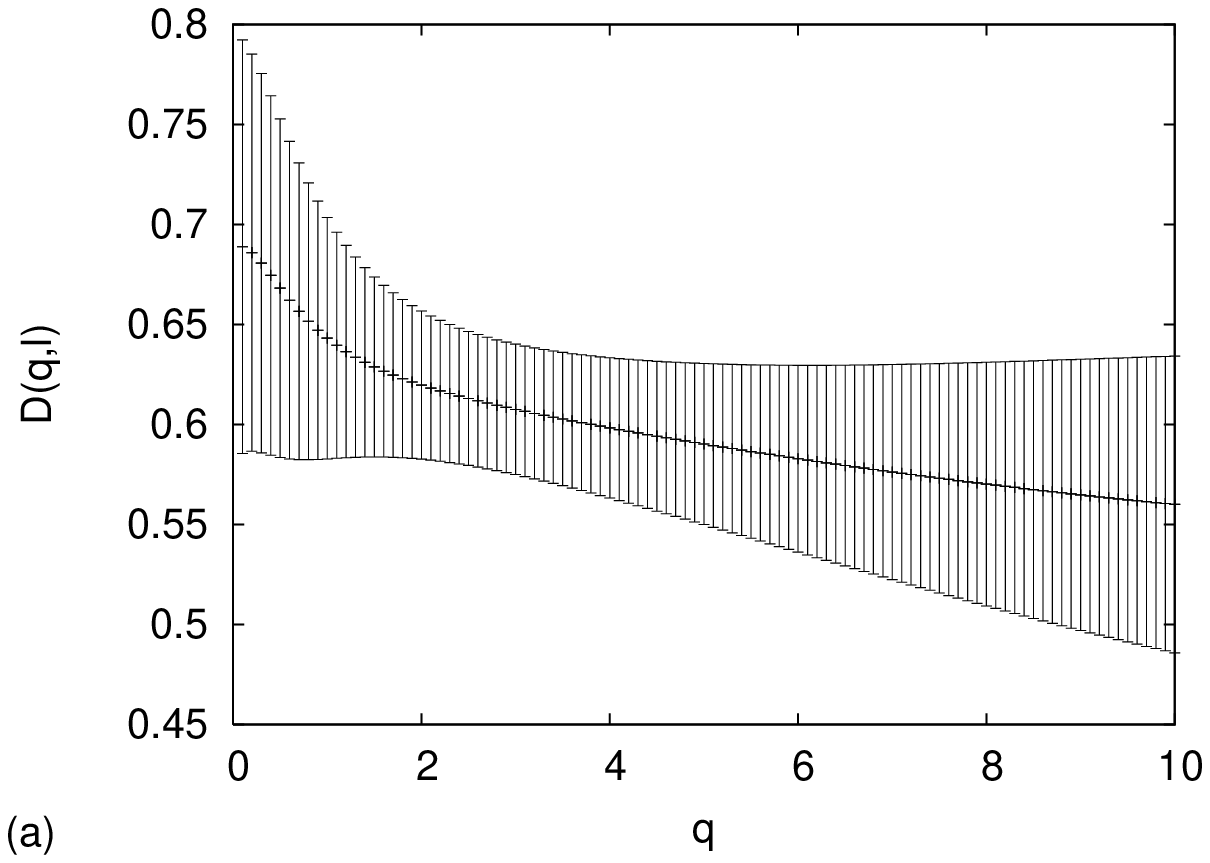}
\includegraphics[width=0.48\textwidth]{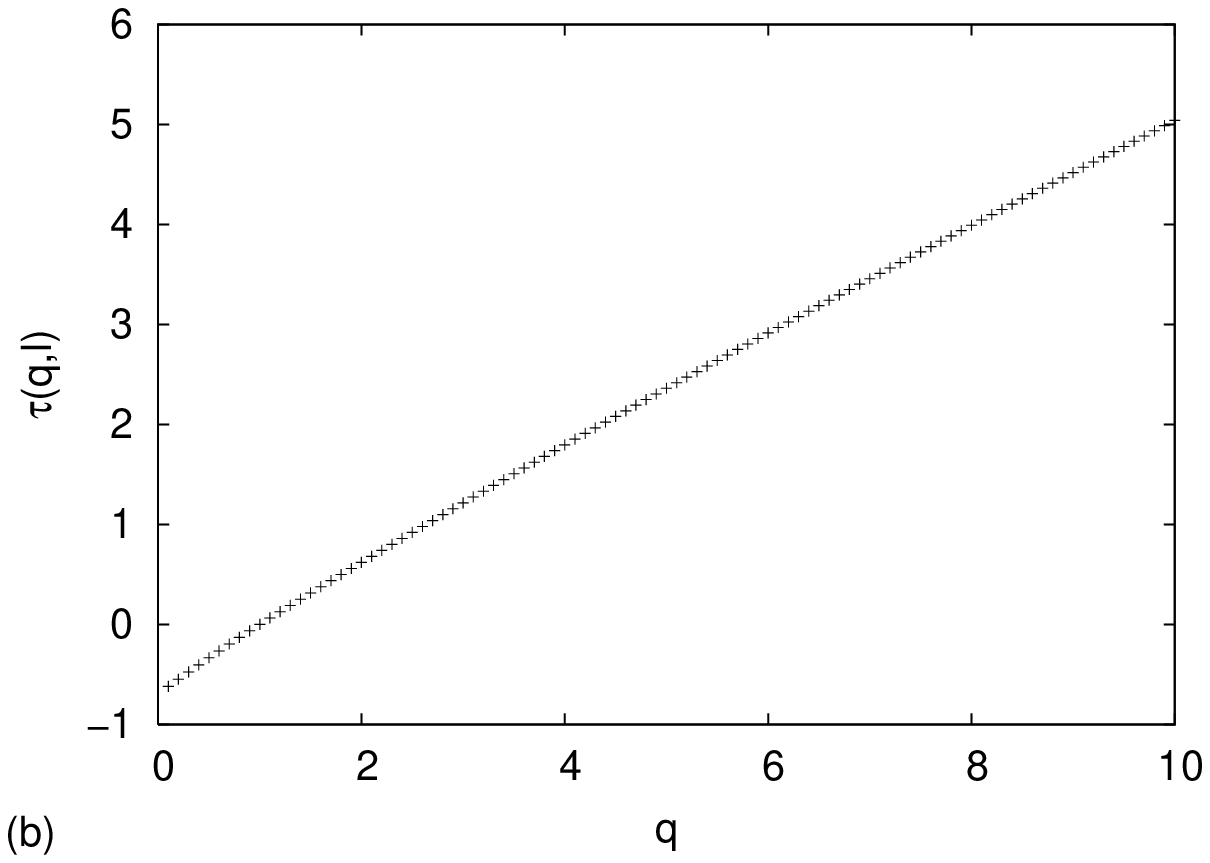}

\caption{\label{fig7} Generalized dimension $D_{q}$ vs. $q$ (panel a) and
global scaling index $\tau_{q}$vs. $q$ (panel b) in configuration
($x)$space for the quintic model at T=14 with $q>0.$.}
\end{figure}

\begin{figure}[!ht]
 \includegraphics[width=0.48\textwidth]{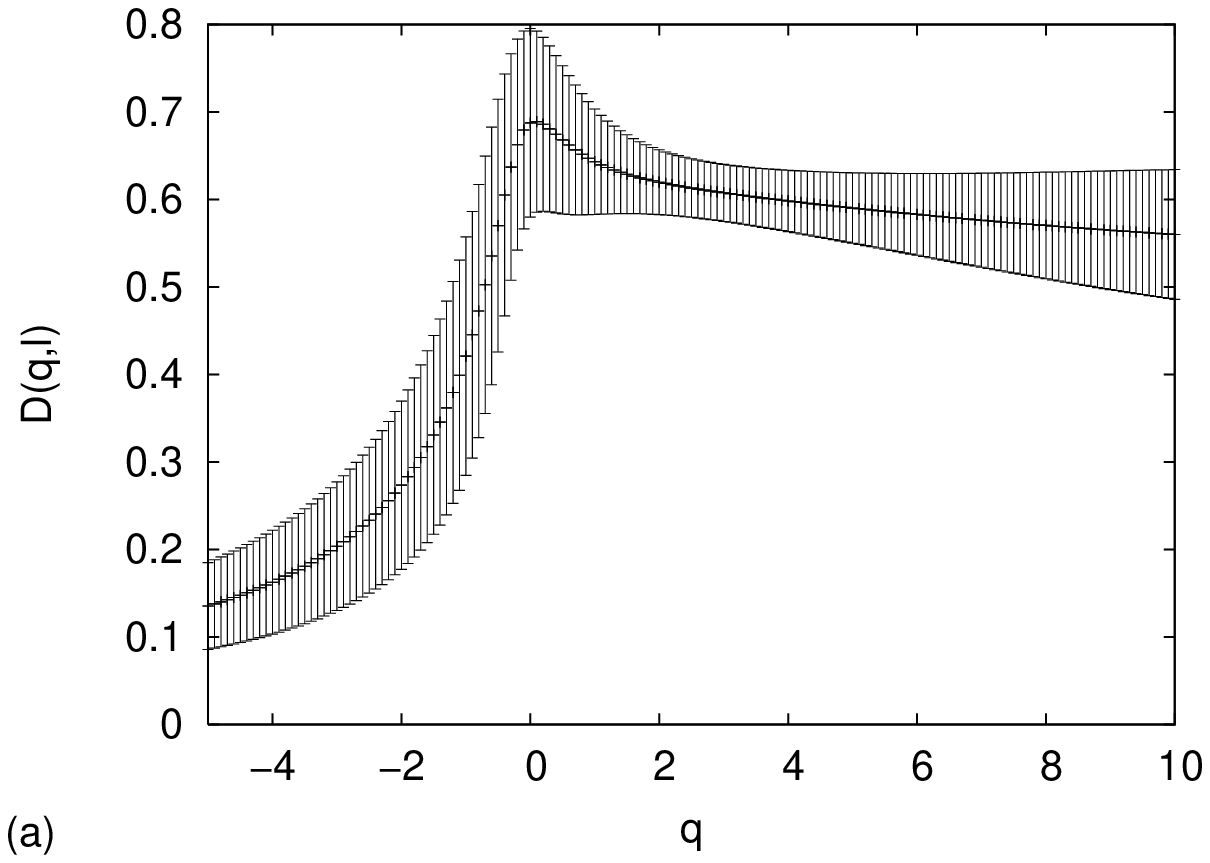}
\includegraphics[width=0.48\textwidth]{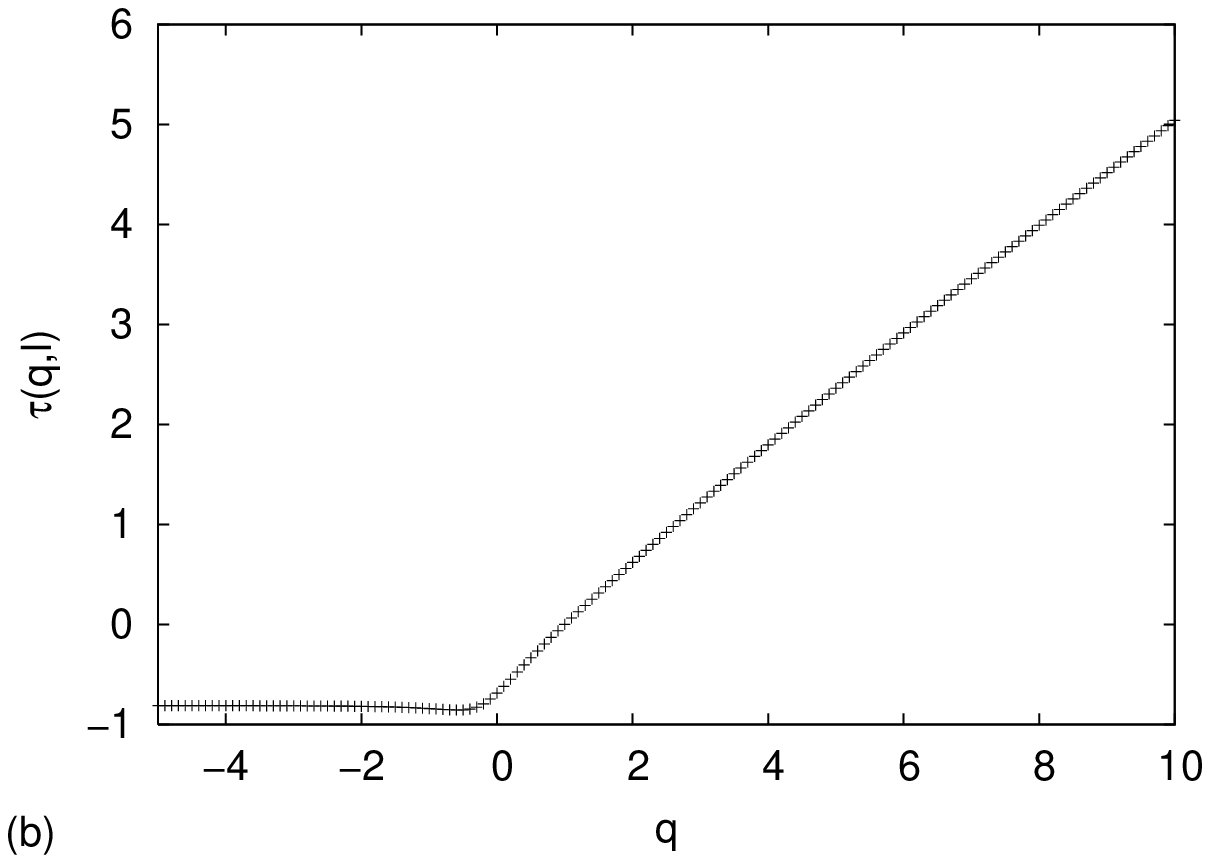}

\caption{\label{fig8} Generalized dimension $D_{q}$ vs. $q$ (panel a) and
global scaling index $\tau_{q}$vs. $q$ (panel b) in configuration
($x)$space for the quintic model at T=14 with $q<0.$.}
\end{figure}

Now that we have identified the important scales, we are able to compute
the generalized dimension, $D_{q}$, and the global index, $\tau_{q}$.
In Fig. 5(a) we plot $D_{q}$ vs $q$ calculated in the $\mu$-space
for the quintic model at $T=14$ for $q>0$ for the case of the larger
scaling range. As expected, it is a decreasing function and clearly
demonstrates multifractal behavior. Although $\mu$ space is two dimensional,
$D_{0}$ is about 0.9. While $D_{0}>D_{1}$ is typical for multifractals
(information dimension covers the important region), the fact that
$D_{q}$ is strictly decreasing suggests greater \char`\"{}fractality\char`\"{}
from increasingly overdense regions, i.e. the system is strongly inhomogeneous.
In Fig. 5b we plot $\tau_{q}$ versus $q$ in mu space. Two linear
regions ($0<q<1$ and $1<q<10$) can be distinguished suggesting bifractal
geometry, i.e. a superposition of two fractals with unique values
of $\alpha$ and $f(\alpha)$. 

We have seen that the one dimensional gravitational system reveals
fractal geometry in the higher dimensional $\mu$ space (frequently
referred to as phase space in the astronomical literature). Historically
self similar behavior was first inferred in gravitational systems
by studying the distribution of galaxy locations, i.e in the system's
configuration space, and searching for powerlaw behavior in the correlation
function (see above).\cite{bs} In common with this approach, we project
the mu space distribution on the (position) line to study the geometry
of the configuration space. In Fig. \ref{fig6} a,b,c,d, we provide
plots of $1/(q-1)ln(I_{q})$ vs $ln(l)$ in $x$ (configuration) space
for the quintic model prepared as above at T=14. In common with the
$\mu$-space distribution, for $q=-5$ and $q=0$, there is a single,
large scaling region ($-3<ln(l)<3$ for $q=-5$ and $-2<ln(l)<6$
for $q=0$). Similarly, at $q=5$ (c) and $q=10$ (d) the existence
of two scaling ranges becomes apparent, $-4<ln(l)<0$ and $0<ln(l)<8$
for $q=5$ (about 3.3 decades), $-3<ln(l)<1$ and $1<ln(l)<8$ for
$q=10$ (about 3 decades). Note that the scaling ranges are similar
for both the $\mu$ and configuration space, but they are not identical.

In Figure \ref{fig7} we illustrate the behavior of the generalized
dimension $D_{q}$and global index $\tau_{q}$ for the configuration
space of the quintic model at $T=14$. Although now the embedding
dimension is $d=1$, $D_{0}$ is about $0.7$ so the distribution
is definitely fractal. As anticipated from the $\mu$-space distribution,
$D_{q}$ is a decreasing function of $q$ so it is also inhomogeneous
and multifractal. In Fig. \ref{fig7}b we plot $\tau_{q}$ vs $q$
in configuration space, for the same system. In contrast with the
$\mu$-space index, here we observe a nearly linear function with
slight convexity (decreasing derivative with respect to $q$). Perhaps
this results from the superposition of three linear regions with distinct
$\alpha$ and $f(\alpha)$, say with $0<q<1.5$; $1.5<q<7$; $7<q<10$,
but this cannot be inferred from the plot without further analysis.

For completeness, in Fig. \ref{fig8} we examine the behavior of $D_{q}$
and $\tau_{q}$ for $q<0$. In general, negative $q$ is important
for revealing the geometry of low density regions (voids). We see
that $D_{q}$ has a very unphysical appearance - it is increasing.
The source of the problem can be determined by examining the behavior
of $\tau_{q}$. It is nearly constant over the range $-10<q<0$ with
a value of approximately $\tau_{q}\simeq-1$, implying that the underdense
regions are dominated by a single local dimension.

Part of our goal is to compare how fractal geometry arises in a family
of related models. So far we have presented results for the quintic,
or Q, model. However we have also carried out similar studies of the
Rouet-Feix (RF) model, the model without friction (which can be obtained
from either of the former by nullifying the first derivative contribution
in Eq(\ref{final}), and simply an isolated system without friction
or background. The latter is a purely Newtonian model without a cosmological
connection. In Table \ref{tab1} we list the important characteristic dimensions
and exponents for the the quintic model and the model without friction
for comparison. While there are similarities in the fractal structure
of each of these models, they are not the same. For example in the
quintic model the generalized fractal dimension, $D_{q}$, is consistently
smaller in each manifold than the corresponding dimension in the model
without friction. It is noteworth that the box counting dimension
in the configurqtion space of the quintic model is 0.69 compared with
0.90 for the model without friction. Moreover $D_{q}$ is decreasing
more rapidly in the quintic model demonstrating stronger inhomogeneity.

\begin{table}[htdp]
\caption{default}
\begin{center}
\begin{tabular}{c c  c c c c c}
 &\hspace*{.5cm} & \multicolumn{2}{c}{Quintic model} &\hspace*{.5cm} &\multicolumn{2}{c}{RF model} \\ \cline{3-4} \cline{6-7}
$q$ & & x-space          & $\mu$-space       & & x-space            & $\mu$-space   \\
0    & & $0.69\pm0.11$ & $0.89\pm0.06$  & & $0.89\pm0.02$ &   $1.14\pm0.04$ \\                                                 
1    & & $0.64\pm0.06$ & $0.84\pm0.07$  & & $0.88\pm0.04$ &   $1.17\pm0.04$ \\                       
2    & & $0.62\pm0.04$ & $0.81\pm0.05$  & & $0.85\pm0.04$ &   $1.13\pm0.05$  \\                       
\end{tabular}
\end{center}
\label{tab1}
\end{table}%

\begin{figure*}[!ht]
 \centerline{\includegraphics[width=1\textwidth]{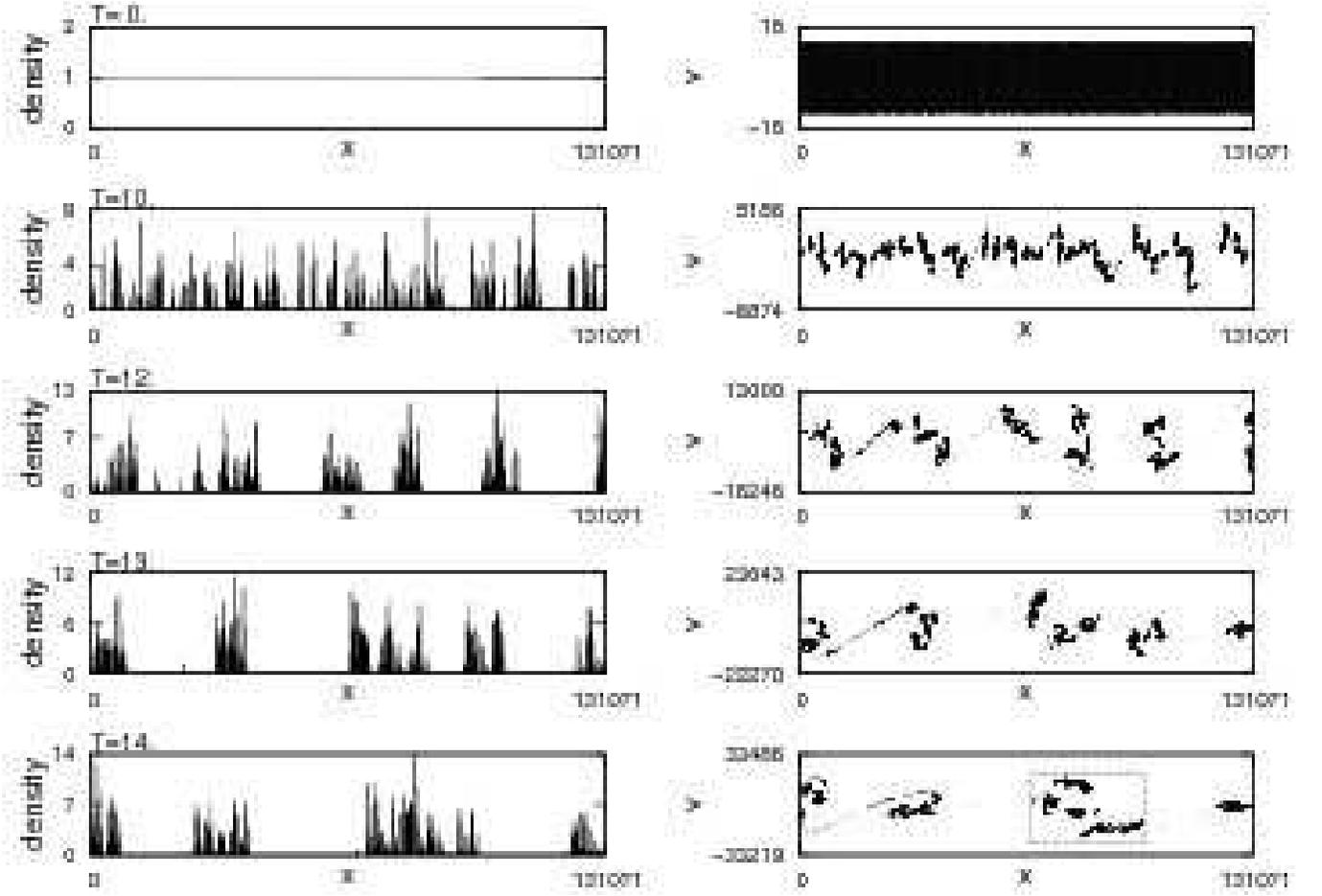}}

\caption{\label{fig9} Evolution in configuration and space for the model
without friction with $2^{17}$ particles from T=0 to T=14. The initial
distribution is a waterbag with velocities in the range (-14, 14)
in the dimensionless units employed here and a size of about 10,400
Jeans' lengths.}
\end{figure*}

\begin{figure}[!ht]
 \includegraphics[width=0.48\textwidth]{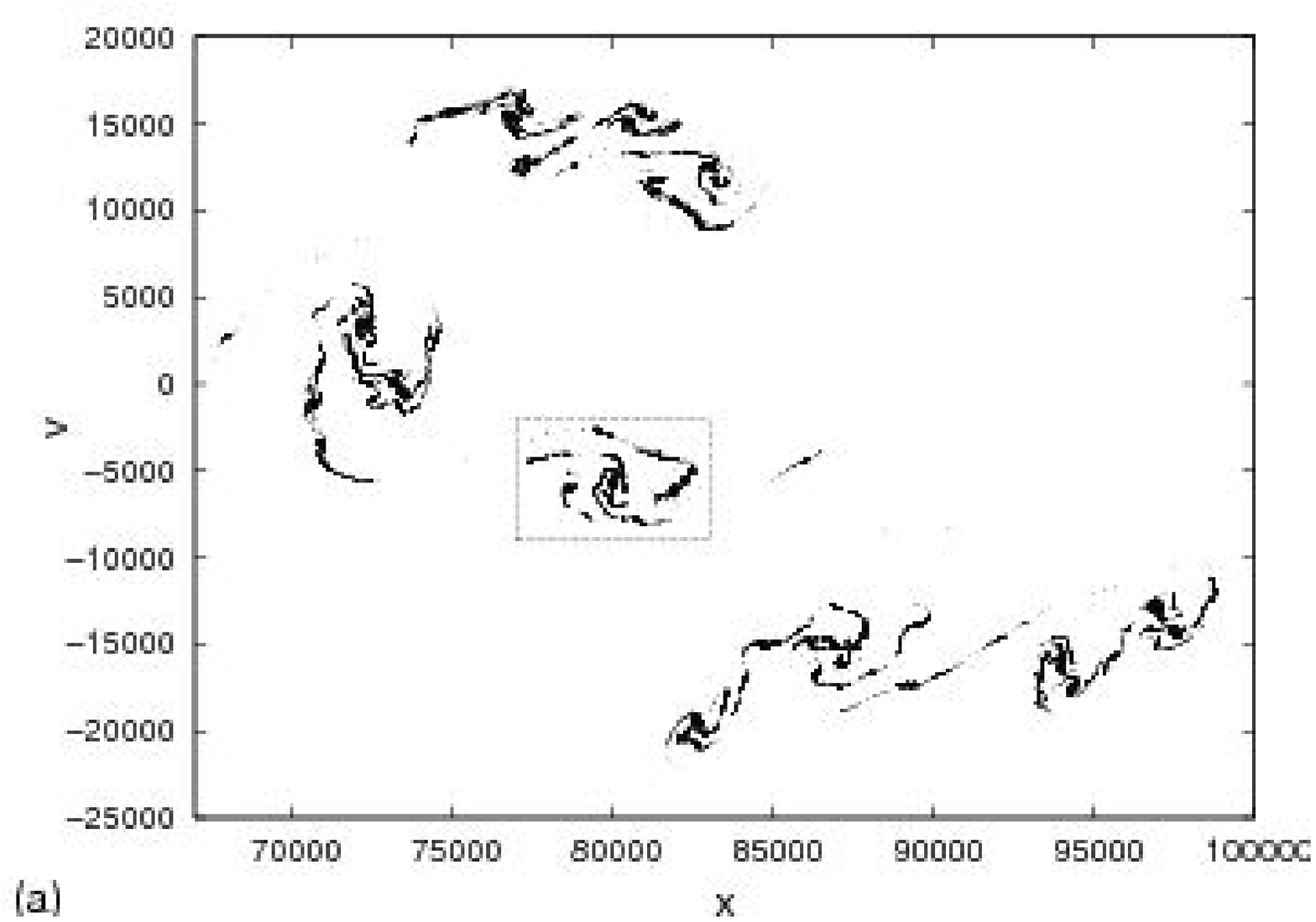}
\includegraphics[width=0.48\textwidth]{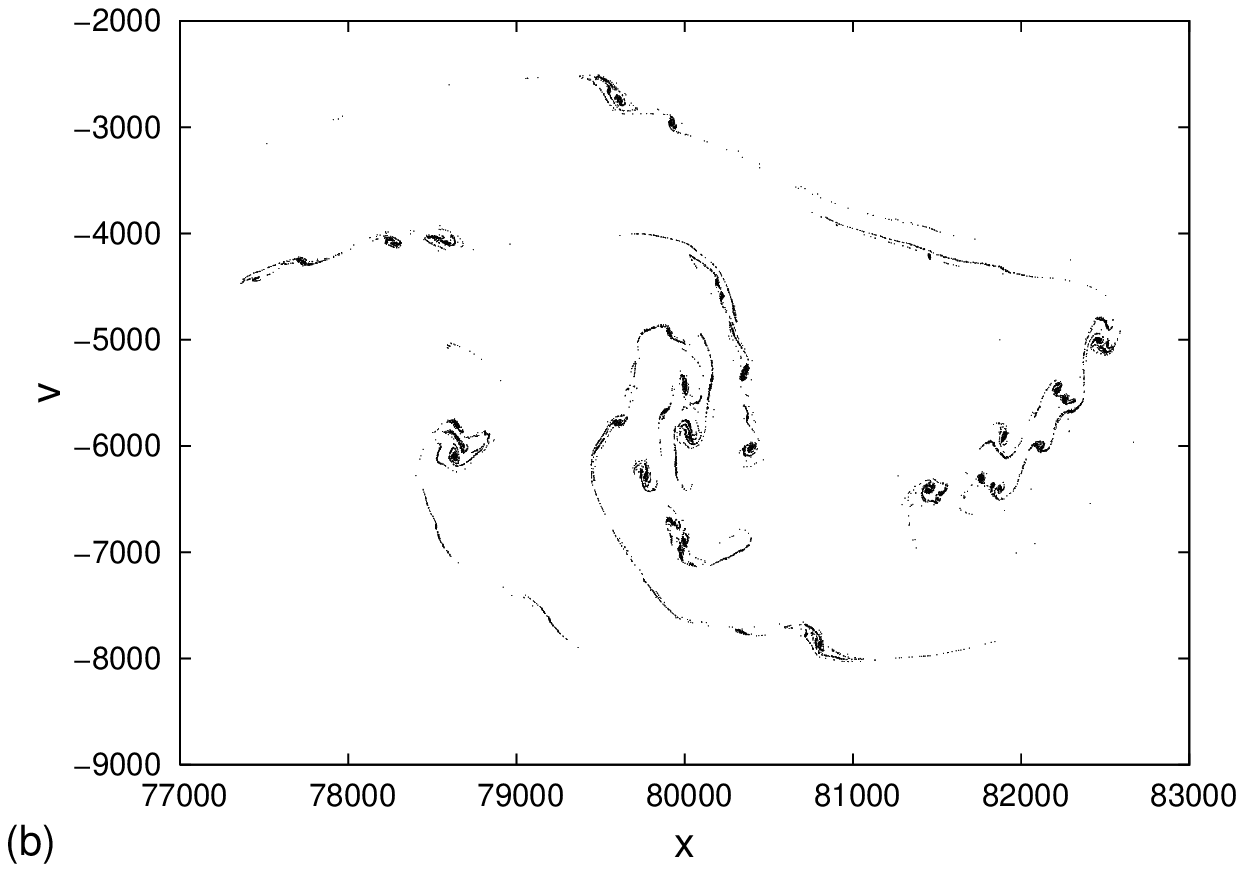}

\caption{\label{fig10} Consecutive expansions (zooms) of successive small
squares (square within a square) selected in the $\mu$ space panels
at T=14. They also have the appearance of a random fractal which suggests
self-similarity.}
\end{figure}

\begin{figure}[!ht]
 \includegraphics[width=0.48\textwidth]{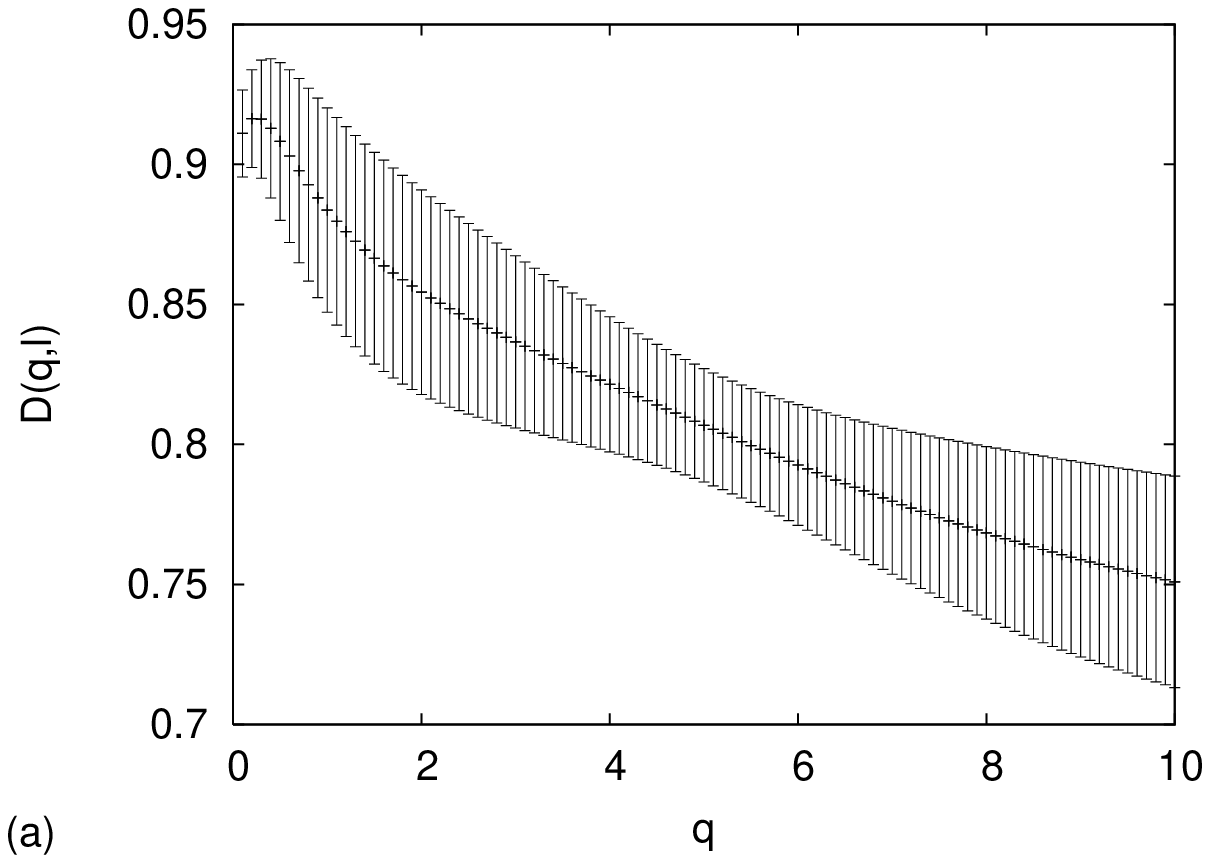}
\includegraphics[width=0.48\textwidth]{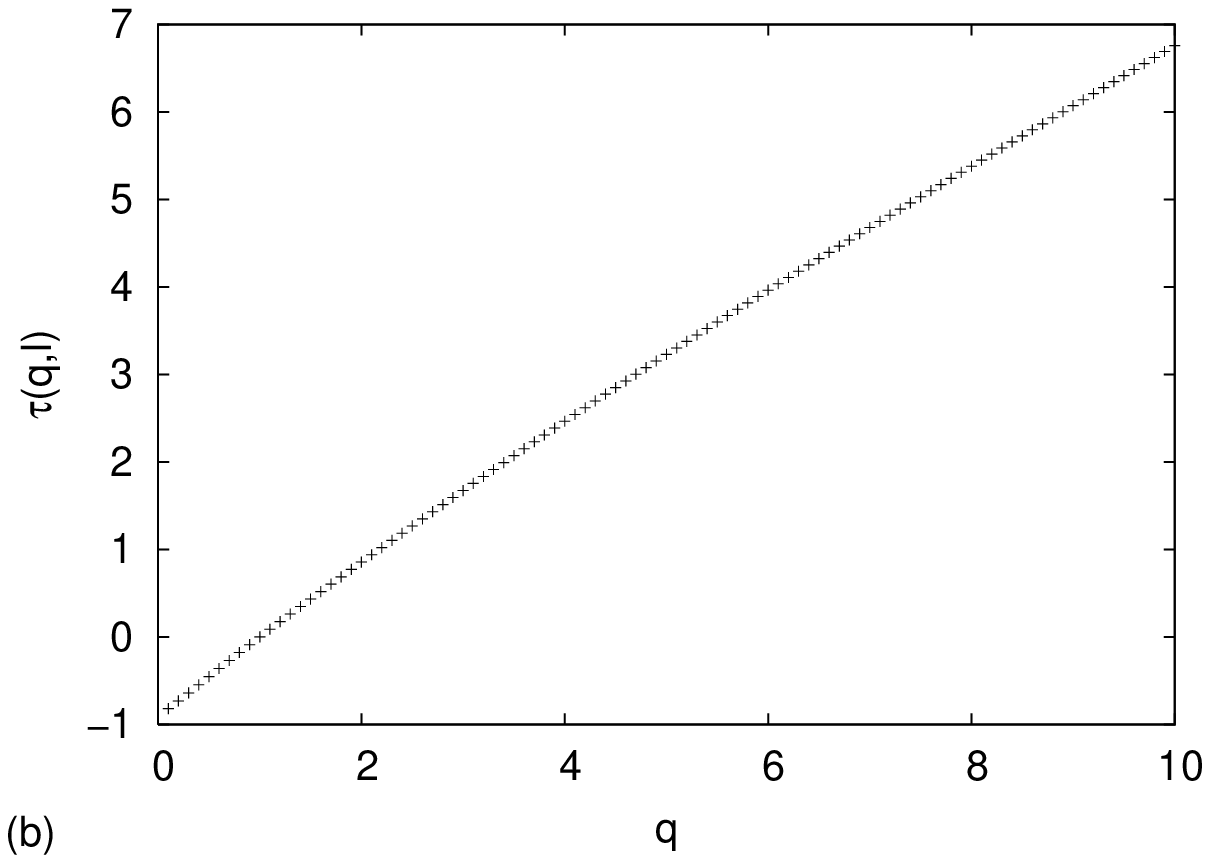}

\caption{\label{fig11} Generalized dimension $D_{q}$ vs. $q$ (panel a)
and global scaling index $\tau_{q}$vs. $q$ (panel b) in configuration
($x)$space for the quintic model at T=14 with $q<0.$}
\end{figure}

The differences in dynamics between the systems can be seen clearly
in Fig. \ref{fig9} where we show the evolution in configuration and
$\mu$-space of the system without friction. In common with the simulations
of the quintic model presented above, the initial conditions are a
waterbag with velocities sampled independently and confined to ($-12.5,12.5$)
in the dimensionless units defined earlier. As before, there are $2^{17}$
particles and the simulation spans $14$ units of the scaled time.
The initial size encompasses about 10,400 Jeans lengths. Compared
with the quintic model, note that here larger clusters form earlier
and have a qualitatively different shape. Due to the absence of {}``friction''
in the dynamics, the velocity spread is larger and there are fewer
clusters at each epoch. At T=12 and T=13 the linear structure of the
underdense regions (voids) in $\mu$ space connecting the clusters
is apparent. Note that they all share a common slope which is due
to stretching in the phase plane. However, equation \ref{stretching}
is no longer adequate to provide the line slope. In Fig. \ref{fig10}
we examine the consecutive expansions (zooms) of the large cluster
on the right hand side of the $\mu$ space distribution at $T=14$.
Once again there is qualitative evidence of stochastic self-similarity
\cite{Ott} which requires analysis for confirmation. To save space,
here we don't reproduce the plots of $\frac{1}{q-1}{\ln(C_{q})}$
versus $\ln{l}$ in each space. Qualitatively they are similar to
the quintic case, but the scaling ranges are less robust.

The behavior of $D_{q}$ and $\tau_{q}$ are similar to that found
for the quintic model. However, because the scaling ranges are less
robust, there is more noise. For example, in Fig. \ref{fig11}a,b.
we plot $D_{q}$ and $\tau_{q}$ versus $q$ in configuration space
for the model without friction. Note the similarities with Fig. \ref{fig6}
for the quintic model. The slope of $\tau_{q}$ gradually decreases
with increasing $q$. There appears to be three linear segments, from
(0, 3), (3, 5), and (5,10) suggesting contributions from three possible
fractal scaling regions.

\section{\label{dis}Discussion of Results}

As mentioned earlier, it is well established that, for a regular multifractal,
the generalized dimension, $D_{q}$ , is a decreasing function of
its argument.\cite{Fed}\cite{Ott} Therefore, for $q<0$ , it would
be incorrect to interpret the simulation results as true generalized
dimensions. There must be an alternative explanation for the behavior
we observe. The picture for positive $q$ is rather different. The
two scaling regions give completely different results. Although one
would suspect from the definition of the generalized dimension that
the scaling region with smaller $l$ would give the correct one, this
is hard to accept since typical plots of the function $D_{q}$ are
still increasing until about $q=2$ (not shown) for this range. On
the other hand, the second, larger, scaling region manifests a well
behaved decreasing function which appears to approach a constant value
of $D_{q}\cong0.63$ for $q\lesssim10$.

It is interesting that we have observed similar behavior with a well
characterized, textbook, fractal that is discussed in numerous sources.\cite{Fed}
As a test of our computational approach we simulated the multiplicative
binomial process. For this multifractal ${\tau_{q}}$ and $D_{q}$
are known precisely.\cite{Fed} When we carried out the fractal analysis
using the methods described above, we also found two scaling regions.
What is most striking is that the scaling region with larger values
of $l$ yielded a ${\tau_{q}}$ (and therefore $D_{q}$) which agreed
to within numerical error with the theoretical prediction! This type
of behavior has been manifested in other simulations of fractal sets.
The formation of the smaller scaling range has been attributed to
the presence of noise in the data. The existence of a second scaling
range has been rigorously shown to arise when Gaussian noise is added
to a standard fractal.\cite{frac_noise} In the simulation of the
multiplicative binomial process, the source of the noise is the random
location of the data points within the smallest bins. At this time
the source of the apparent noise in the one dimensional gravitational
simulations is not precisely known. While it may arise simply from
numerical considerations, there are alternative possible explanations.
For example, noise may arise from sub-Jeans length fluctuations in
the initial data, or from other small scale features of the initial
state. In addition to the box-counting approach, other methods based
on the correlation function formalism that employ the point-wise dimension
can also be used to investigate the fractal properties of the system.
It was also used in our investigation and gives similar results to
the box counting method reported here.

For $q>0$ we appear to be seeing a similar phenomena to the mutiplicative
binomial process or the systems with injected noise.\cite{frac_noise}
Then how do we explain the surprising and counterintuitive results
for $q<0$? Since for negative $q$ we obtain a nearly constant value
for $\ {\tau_{q}}$ from each scaling region, it seems safe to assume
that a region of the data characterized by a simple fractal behavior
has the dominant influence. Moreover, since it only involves $q<0$,
it represents the regions of low density, i.e. the voids. Referring
to the multifractal formalism, we pointed out earlier that in a region
where a single structure dominates,\begin{equation}
\tau_{q}=\alpha q-f(\alpha),\end{equation}
 where $\alpha$ is the strength of the local singularity, or the
local pointwise dimension, and $f(\alpha)$ is the Hausdorf dimension
of its support.\cite{Fed}\cite{Ott} Then the computations show that
for $q<0$ we must have $\alpha=0$ and $f\cong0.9$. This suggests
that the results for negative $q$ are dominated by regions of such
low density that widely separated, ''isolated'', particles are responsible
for the spurious behavior of $D_{q}$. At this time this is simply
a conjecture which needs to be investigated with further computation.

Finally let's reconsider the plot of ${\tau_{q}}$ versus $q$ (Fig.
\ref{fig5}b) for the larger scaling region with $q>0$. As we mentioned
earlier, for $0<q<1.5$ and for $2<q<10$ the curve appears linear
with different slopes in each region. This may be the manifestation
of bifractality first discussed by Bailin and Shaffer for galaxy positions.
The first interval may represent the true fractal structure of the
under-dense regions, while the dense clusters are dominant for the
larger $q$ values. Bifractal behavior is not unique to gravitational
systems. It occurs in a number of well studied model systems, for
example as a superposition of two Cantor sets,\cite{frac_prod} or
from the truncation of Levy flights \cite{levy}\cite{Nakao} .

\section{Conclusions}

\label{con} We have seen that one dimensional models develop hierarchical
structure and manifest robust scaling behavior over particular length
and time scales. In addition, for a number of reasons, they have an
enormous computational advantage over higher dimensional models. First,
it is possible to study the evolution of large systems with on the
order of $2^{17}$particles per dimension. Second, in contrast with
three dimensional N-body simulations, the evolution can be followed
for long times without compromising the dynamics. In particular, the
force is always represented with the accuracy of the computer - there
is no softening. As a consequence it is possible to investigate the
prospect for fractal geometry with some confidence.

In common with 3+1 dimensional cosmologies, with the inclusion of
the Hubble expansion and the transformation to the comoving frame,
both the 1+1 dimensional Quintic and RF \cite{MRexp} models reveal
the formation of dense clusters and voids. They also show evidence
for bifractal geometry. This may be a consequence of dynamical instability
that results in the separation of the system into regions of high
and low density. An interesting observation is that the lower bound
of the length scale that supports the trivial space dimension of unity
in the configuration space grows with time. To the extent that similar
behavior occurs in the 3+1 dimensional universe, this lends support
for the standard cosmological model on sufficiently large scales.
It also suggests that the scale size for homogeneity will grow with
time. Eventually this may be testable with observation.

We have seen that the system shows evidence of two nontrivial scaling
regions. The type of anomalous behavior of $D_{q}$ and $\tau_{q}$
in the scaling region with a finer partition was also found in the
standard multiplicative binomial process with a similar sample size.
Even with the large sample size employed here, this would suggest
that it is a finite size effect. In that respect it will be interesting
to perform simulations dealing directly with the distribution function
given by the Vlasov equation. Then the low density region will be
well described. This may also be true for the region of negative $q$
where a fractal analysis forces us to conclude that the point-wise
dimension vanishes. Computations with different boundary conditions
reveal similar behavior, but this work is only in the preliminary
stages. Future work will include the investigation of the influence
of correlation in the initial conditions, the consideration of initial
conditions of the type employed in current cosmological N-body simulations
\cite{Virgo}, and the study of the fractal geometry of the under
and over dense regions. We plan to compare out findings with three
dimensional studies of the distribution of voids and halos \cite{Gaite}.

In addition to these important features, the one dimensional simulations
unequivocally demonstrate that the essential structure formation is
taking place in the $\mu$ space (phase space for astronomers). Thus
the apparent fractal geometry that we observe in configuration space
is simply a shadow (projection) of the higher dimensional structure.
The development of structure in $\mu$ space and its apparent fractal
geometry may be the most significant result of these studies. Since
evolution in the six dimensional $\mu$ space of the observable universe
is beyond our current capability, the study of lower dimensional models
is an important guide for understanding the important features of
the higher dimensional evolution. For example, we have shown analytically
that dynamical {}``stretching'' of the $\mu$ space geometry is
responsible for the formation of underdense regions (voids) and, consequently,
the concentration of mass in regions of decreasing area. We are currently
extending this analysis to the study of structure formation in the
more realistic 3+1 dimensional manifold.

\section*{Acknowledgements}

A sabbatical leave from Texas Christian University and the hospitality
and support from the Universite d'Orleans is gratefully acknowledged
by B. N. Miller.

\bibliography{gravbib}

\end{document}